\RequirePackage{fix-cm}
\documentclass[smallextended]{svjour3}       
\smartqed  
\usepackage{amssymb}
\usepackage{caption}
\usepackage{url}
\usepackage{graphicx}
\usepackage{amsmath}
\usepackage{float}
\usepackage{caption}
\usepackage{color}
\usepackage[mathscr]{euscript}
\usepackage{euscript}

\graphicspath{{./Images/}}

\def \f{\vec{f}}

\def\f {{{\bf f}}}
\def\u {{{\bf u}}}

\def\div {{\nabla \cdot}}

\begin{document}

\title{Don't be jelly:
}
\subtitle{Exploring effective jellyfish locomotion}


\author{Jason G. Miles         \and
        Nicholas A. Battista 
}


\institute{J. G. Miles \at
              Dept. of Mathematics and Statistics\\
              The College of New Jersey, Ewing Township, NJ, USA \\
              \email{milesj4@tcnj.edu}           
           \and
           N. A. Battista \at
              Dept. of Mathematics and Statistics\\
              The College of New Jersey, Ewing Township, NJ, USA \\
              \email{battistn@tcnj.edu}           
}

\date{Received: date / Accepted: date}

\maketitle

%
%
%
%
%
%

\begin{abstract}

Jellyfish have been called one of the most energy-efficient animals in the world due to the ease in which they move through their fluid environment, by product of their morphological, muscular, and material properties. We investigated jellyfish locomotion by conducting \textit{in silico} comparative studies and explored swimming performance across different fluid scales (e.g., Reynolds Number), bell contraction frequencies, and contraction phase kinematics for a jellyfish with a fineness ratio of 1 (ratio of bell height to bell diameter). To study these relationships, an open source implementation of the immersed boundary method was used (\textit{IB2d}) to solve the fully coupled fluid-structure interaction problem of a flexible jellyfish bell in a viscous fluid. Thorough 2D parameter subspace explorations illustrated optimal parameter combinations in which give rise to enhanced swimming performance. All performance metrics indicated a higher sensitivity to bell actuation frequency than fluid scale or contraction phase kinematics, via Sobol sensitivity analysis, on a high performance parameter subspace. Moreover, Pareto-like fronts were identified in the overall performance space involving the cost of transport and forward swimming speed. Patterns emerged within these performance spaces when highlighting different parameter regions, which complemented the global sensitivity results. Lastly, an open source computational model for jellyfish locomotion is offered to the science community that can be used as a starting place for future numerical experimentation.


\end{abstract}

%
%
%
%

\section{Introduction}

Scientists have long tried to understand aquatic locomotion in organisms ranging in size and scale from phytoplankton to whales. Oftentimes, an organism's size and shape dictates what type of locomotive process it uses to move around its fluid environment (e.g., small invertebrates such as ctenophores use ciliary based motion \cite{Pang:2008}, while larger organisms like fish may use fin-based propulsion \cite{Akanyeti:2017}). One quantity that helps describe the \textit{scale} of a fluid system is a dimensionless number called the Reynolds Number, $\mathrm{Re}$. The $\mathrm{Re}$ can be interpreted as the ratio of inertial to viscous forces, thus categorizing the importance of viscous effects on the behavior of the underlying fluid system. It can be mathematically represented by
\begin{equation}
    \label{eq:Re} \mathrm{Re} = \frac{\rho VL}{\mu} = \frac{\rho f L^2}{\mu},
\end{equation}
where $\rho$ and $\mu$ are the fluid's density and viscosity, respectively, $L$ is a characteristic length scale of the problem (e.g., such as the size of a fish fin or ctenophore's cilia), and $V=fL$ is a characteristic velocity scale, which could be the speed of a fish fin or ctenophore's cilia during it's stroke cycle. Although in their natural environments a salt water fish and ctenophore are immersed within approximately the same fluid (e.g., same density and viscosity), they use considerably different modes of locomotion (mechanisms) to move through their fluid environment. There are evolutionary processes which selected and developed particular locomotive mechanisms. Many such mechanisms are largely based on the size of the organism, e.g., length scale of the problem \cite{Vogel:1996}. 

Recently, scientists have been trying to understand how jellyfish, the most energy-efficient animals in the world \cite{Gemmell:2013,Costello:2020}, swim. Jellyfish are soft body marine organisms composed of gelatinous bell, tentacles containing nematocists for prey capture, and either 4 or 8 oral arms \cite{Hamlet:2012,Santhanakrishnan:2012}. Their nervous system typically consists of a distributed net of cells, which are concentrated into small structures called rhopalia \cite{Gemmell:2014}. There are between four and sixteen rhopalia around the rim of the bell, which coordinate muscular contraction to propel the jellyfish forward \cite{Satterlie:2011}. Their relatively simple morphology and nervous systems make them attractive to robotocists \cite{Joshi:2012,Frame:2018,Christianso:2019,Ren:2019,Costello:2020}. In 2020, a bio-hybrid robot was introduced that could swim up to 2.8 times faster than the natural swimming speeds of the actual jellyfish to which the microelectronics were planted. It was able to do this while requiring 10-1000 times less power per unit mass than other aquatic robots \cite{Xu:2020}.

Outside of laboratory settings, many computational scientists have developed computational fluid dynamics (CFD) models of jellyfish that produce forward propulsion \cite{Dular:2009,Wilson:2009,Sahin:2009a,Sahin:2009b,Lipinski:2009,Rudolf:2010,Hershlag:2011,Wilson:2011,Alben:2013,Gemmell:2013,Park:2014,Yuan:2014,Park:2015,Hoover:2015,Hoover:2017,Hoover:2019,Miles:2019b,Pallasdies:2019} and have compared swimming performance over a large mechanospace of bell flexibility, muscular contraction strength, and contraction kinematics. Computational studies are attractive to scientists as it is easier (e.g., more cost and time efficient) to do parameter studies using computational models rather than building many robots or physical models. Some scientists have begun producing robust multi-scale models, whether in 2D or 3D, that couple the underlying electrophysiology to muscular force generation that results in propulsion forward \cite{Hoover:2017,Hoover:2019,Pallasdies:2019}. While these models incorporate a multitude of biological data, they inherently come with the expense of more modeling degrees of freedom, i.e., parameters. 

On the other hand, there is still much to be learned about jet propulsion from simple 2D CFD actuator models of jellyfish. Hoover et al. 2015 \cite{Hoover:2015} demonstrated enhanced swimming performance when bells were driven at their resonant frequencies. Peng and Alben 2012 \cite{Peng:2012} and Alben et al. 2013 \cite{Alben:2013} illustrated how variations in bell contraction kinematics could result in substantially different swimming behavior. However, these studies did not investigate performance across different fluid scales ($\mathrm{Re}$) or extensively explore swimming performance across large parameter subspaces. Studying these mechanisms across different scales could provide more context into the behavior of some jellyfish, to whom exhibit a jet propulsion mode of locomotion only during development. After maturing (when their bells grow wide enough), they switch to a more efficient swimming mode - jet-paddling or rowing \cite{McHenry:2003,Weston:2009,Blough:2011}. Furthermore, 2D models of prolate jellyfish (the jellyfish who use jet propulsion) have shown both qualitative and quantitative agreement with experimental data \cite{Sahin:2009b,Hershlag:2011,Alben:2013,Kakani:2013}, making it more feasible to perform widespread parameter studies than their 3D computational model counterparts.

In this work, we used an open-source implementation of the immersed boundary method, \textit{IB2d} \cite{Battista:2015,BattistaIB2d:2017,BattistaIB2d:2018}, to model jellyfish locomotion using a fully coupled fluid-structure interaction framework in 2D. In particular, we modified the simple actuator model from Hoover et al. 2015 \cite{Hoover:2015} and performed comparative studies across different fluid scales (i.e., varying Reynolds Number, $\mathrm{Re}$), bell contraction frequencies ($f$), and contraction (and expansion) phase kinematics (quantified by a kinematic parameter, $p$). The parameter $p$ gives the fraction of the overall bell actuation cycle in which the bell was actively contracting. Thus, we varied three parameters ($\mathrm{Re},f,p$) and thoroughly (\textit{intensely}) explored 2D parameter subspaces consisting of ($\mathrm{Re},f$), ($\mathrm{Re},p$), and ($f,p$). Each 2D subspace was explored three times, each corresponding to a different value of the third free parameter. Furthermore, global sensitivity analysis was performed via Sobol sensitivity analysis to isolate which parameter(s) most significantly affected jellyfish swimming performance, as first posited by Dabiri \& Gharib 2003 \cite{Dabiri:2003}, but had yet to have been quantified. A model is called \textit{sensitive} to an input parameter if small variations in the parameter result in great changes in the system's output. Global sensitivity analyses attempt to quantify the impact of variations among the input parameters on the overall model output(s) in a holistic fashion \cite{Saltelli:2002}, as opposed to only local sensitivity measures \cite{Link:2018,Saltelli:2019}. The resulting performance data from the Sobol analysis was projected onto 2D subspaces for comparative purposes. Lastly, Pareto optimal-like fronts were identified across the performance space between cost of transport and forward swimming speed \cite{Eloy:2013,Verma:2017,Smits:2019}. These data exhibited qualitative agreement with the sensitivity results. 

In addition, we offer the science community the first open-source jellyfish locomotion model in a fluid-structure interaction framework. It can be found at \url{github.com/nickabattista/IB2d} in the sub-directory: 

\begin{center}
\texttt{IB2d$/$matIB2d$/$Examples$/$Examples$\_$Jellyfish$\_$Swimming$/$Hoover$\_$Jellyfish}.
\end{center}

%
%
%
%

\section{Mathematical Methods}
\label{sec:methods}

To model a flexible jellyfish bell immersed in a viscous, incompressible fluid, we used a fluid-structure interaction (FSI) framework. In particular, we used an implementation of the immersed boundary method (IB), which was first conceived in the 1970s by Charles Peskin \cite{Peskin:1972,Peskin:1977,Peskin:2002}. The immersed boundary method has since been improved upon numerous times \cite{Fauci:1993,Lai:2000,Cortez:2000,Griffith:2005,Mittal:2005,Griffith:2007,BGriffithIBAMR,Griffith:IBFE} and is still a leading numerical framework for studying problems in FSI due to its robustness, simplicity, and flexibility in modelling complex deformable structures, like many that arise in biological contexts \cite{BattistaIB2d:2017,BattistaIB2d:2018}. 

It has previously been applied to study problems ranging from cardiac fluid dynamics \cite{Miller:2011,Griffith:2012b,Battista:2017,Battista:2020} to aquatic locomotion \cite{Bhalla:2013a,Bhalla:2013b,Hamlet:2015} to insect flight \cite{Miller:2004,Miller:2005,SJones:2015} to parachuting \cite{Kim:2006} to dating and relationships \cite{BattistaIB2d:2017}. Additional details on the IB method are provided in the Appendix \ref{IB_Appendix}.

Below we will discuss the implementation of the jellyfish locomotion model in the open-source \textit{IB2d} framework, i.e., the computational geometry, geometrical and fluid parameters, and model assumptions. Our model is based on the $2D$ jellyfish locomotion model of Hoover et al. 2015 \cite{Hoover:2015} whose model was implemented in the open-source IB software called IBAMR \cite{BGriffithIBAMR}. IBAMR is parallelized IB software with adaptive mesh refinement \cite{MJBerger84,Roma:1999,Griffith:2007}.

%
%

\subsection{Computational Parameters}
\label{sec:parameters}

In this study, we used the frequency-based Reynolds number, $\mathrm{Re}$, to describe the locomotive processes of a jellyfish. The characteristic length, $D_{jelly}$, is set to the bell diameter at rest and the characteristic frequency, $f_{jelly}$, is set to the actuation (stroke) frequency. Therefore our characteristic velocity scale is set to $V_{jelly}=f_{jelly}D_{jelly}$, as in Eq.(\ref{eq:Re}),
\begin{equation}
    \label{eq:Re_study} Re = \frac{\rho f_{jelly} D_{jelly}^2 }{\mu}.
\end{equation}
Fluid parameters (density and dynamic viscosity) can be found in Table \ref{table:num_param}. Across all studies the dynamic viscosity, $\mu$, was selectively chosen to give a specific $\mathrm{Re}$ value, for a particular contraction frequency, $f$. The jellyfish bell's diameter and height remained the same for all simulations performed and were chosen such that the jellyfish's fineness ratio (the ratio of bell height to bell diameter) was exactly equal to $1.0$ \cite{Dabiri:2007}. This allowed us to investigate performance of the jet propulsive locomotion mode directly at the interface between oblate and prolate jellyfish, which have a fineness ratio of less than one and greater than one, respectively. Prolate jellyfish are known to use jet propulsion, while oblate use a jet-paddling propulsion mode \cite{Dabiri:2007}. One such jellyfish with a fineness ratio of $\sim1$ is \textit{Catostylus mosaicus}, the blue blubber jellyfish. However, our study did not explicitly attempt to model this jellyfish due to its complex oral morphology \cite{Neil:2018b}, which accounts for a larger proportion of the jellyfish's mass than other species \cite{Arai:1997}. Previous studies have shown that the addition of tentacles or oral arms can substantially affect swimming performance \cite{Katija:2015,Miles:2019b}. Here we restricted our focus on the locomotion performance via jet propulsion on the bell alone.

\begin{table}
\begin{center}
\begin{tabular}{| c | c | c | c |}
    \hline
    Parameter               & Variable    & Units        & Value \\ \hline
    Domain Size            & $[L_x,L_y]$  & m               &  $[5,12]$             \\ \hline
    Spatial Grid Size      & $dx=dy$      & m               &  $L_x/320=L_y/768$            \\ \hline
    Lagrangian Grid Size    & $ds$        & m               &  $dx/2$               \\ \hline
    Time Step Size          & $dt$        & s               &  $10^{-5}$   \\ \hline
    Total Simulation Time    & $\mathscr{T}$  & \textit{pulses} &  $8$               \\ \hline
    Fluid Density            & $\rho$     & $kg/m^3$        &  $1000$               \\ \hline
    Fluid Dynamic Viscosity & $\mu$       & $kg/(ms)$       &  \textit{varied}      \\ \hline
    Bell Radius           & $a$           & m               &  $0.5$   \\ \hline
    Bell Diameter           & $D$ ($2a$)  & m        &  $1.0$ \\ \hline
    Bell Height           & $h$           & m               &  $1.0$   \\ \hline
    Contraction Frequency   & $f$         & $1/s$          &  \textit{varied}     \\ \hline
    Actuation Period   & $T=1/f$          & $s$          &  \textit{varied}     \\ \hline
    Contraction Phase Fraction & $p$       &           &  \textit{varied}     \\ \hline
    Spring Stiffness   & $k_{spr}$ & $kg\cdot m/s^2$ &  $1\times10^{7}$  \\ \hline
    Beam Stiffness   & $k_{beam} $ & $kg\cdot m/s^2$ &  $2.5\times10^{5}$  \\ \hline
    Muscle Spring Stiffness & $k_{muscle}$ & $kg\cdot m/s^2$ &  $1\times10^{5}$\\
    \hline
    \end{tabular}
    \caption{Numerical parameters used in the two-dimensional simulations.}
    \label{table:num_param}
    \end{center}
\end{table}

For all simulations performed and analyzed, periodic boundary conditions were used on all edges of the computational domain and the computational width was kept constant with $L_x=5.$ Convergence studies demonstrated low relative errors in swimming speeds for domain sizes from $L_x\in[3,8]$ for $Re=\{37.5,75,150,300\}$, see Figure \ref{fig:Conv_Check_speed} in Appendix \ref{app:conv_check}. Amongst all cases, a similar trend was observed where narrower computational domains led to slightly decreased forward swimming speeds while qualitative differences vortex formation were minimal, see Figure \ref{fig:Conv_Check_vortices} in the same appendix. Additional grid resolution convergence studies were performed in Battista \& Mizuhara 2019 \cite{BattistaMizuhara:2019} using the same $2D$ jellyfish computational model that demonstrated appropriate convergence rates in both the Eulerian (fluid) and Lagrangian (jellyfish) data for the numerical parameters described in Table \ref{table:num_param}.

%
%

\subsection{Jellyfish Computational Model}
\label{sec:jelly_model}

\begin{figure}[H]
\centering
\includegraphics[width=0.65\textwidth]{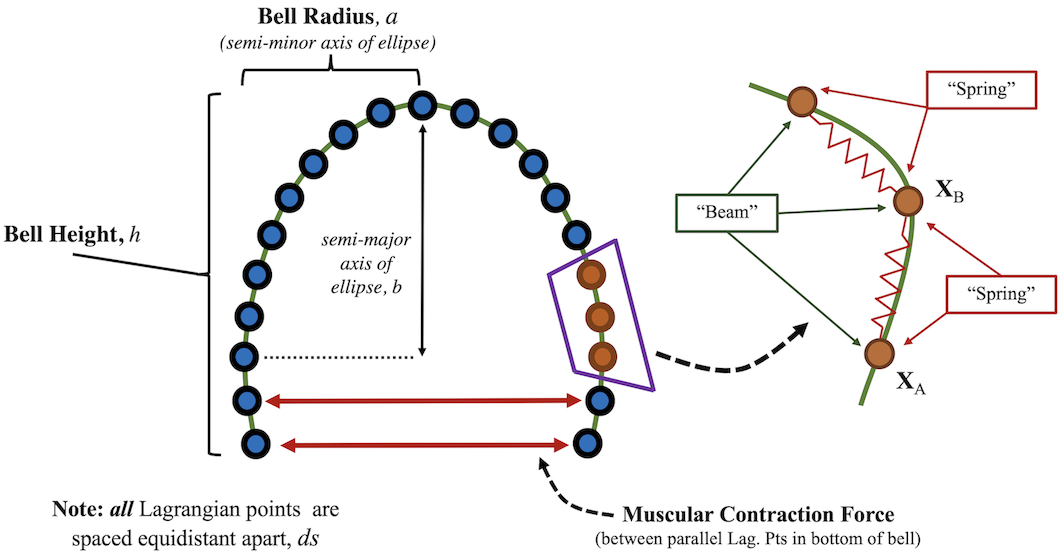}
\caption{Jellyfish model geometry composed of discrete points is a semi-elliptical configuration. The points are connected by virtual springs and virtual beams in the \textit{IB2d}.}
\label{fig:Model_Geometry}
\end{figure}

The geometry was composed of a semi-ellipse of semi-major axis, $b=0.75$, and semi-minor axis, $a=0.5$, see Figure \ref{fig:Model_Geometry}. Note that the overall height of the bell was $h=1.0$, thus the bell was composed of more than only a hemi-ellipse. As shown, it was composed of discrete Lagrangian points that were equally spaced a distance $ds$ apart. We note that this \textit{Lagrangian} mesh was twice as resolved as the background fluid grid, i.e., $ds=0.5dx$ \cite{Peskin:2002}. This choice was made in order to minimize interpolation errors between the fluid and jellyfish grids (see Eqs. (\ref{eq:force1})-(\ref{eq:force2}) in Appendix \ref{IB_Appendix}), which typically manifest themselves as fluid leaking through the immersed structure.  

Although one may view the jellyfish here as being immersed in the fluid, the jellyfish (Lagrangian grid) and fluid (Eulerian mesh), in an IB framework they only through integral equations with delta function kernels (see Eqs. (\ref{eq:force1})-(\ref{eq:force2}) in Appendix \ref{IB_Appendix} for more details.) In a nutshell, the Lagrangian mesh is allowed to move around and change shape. As an observer, one sees the motion of the body. On the Eulerian mesh we only measured what is happening in the fluid (velocity, pressure, external forcing) at discrete rectangular lattice points. In the latter, it is as though we had a number of measurement tools that were only checking the fluid at the precise locations where the tools were placed; we did not track individual fluid blobs. The integral equations with delta function kernels simply say that the fluid points (on the rectangular Eulerian mesh) nearest the jellyfish (on the moving Lagrangian mesh) feel the movement of the jellyfish the most (via a force), as compared to locations in the fluid grid far away (Eq. (\ref{eq:force1})). A similar analogy describes how the jellyfish feels the effect of the fluid motion by the fluid grid points nearest the jellyfish (Eq. (\ref{eq:force2})).

Successive points along the jellyfish were connected by \textit{virtual springs} and \textit{virtual (non-invariant) beams} in the \textit{IB2d} framework, as illustrated in Figure \ref{fig:Model_Geometry}. Virtual springs allowed the geometrical configuration to either stretch or compress, while virtual beams allowed bending between three successive points. When the geometry stretched, compressed, or bent there were elastic deformation forces that arose from the configuration not being in its preferred energy state, i.e., its initial bell configuration. Note that the beams were deemed non-invariant because the preferred configuration is non-invariant under rotations, so if the jellyfish had turned, the model would have undergone unrealistic motion due to these artifacts. 

These deformation forces were computed as below,
\begin{align}
\label{eq:spring} &\text{\footnotesize $\textbf{F}_{spr} = k_{spr} \left( 1 - \frac{R_L(t)}{||{\bf{X}}_{A}(t)-{\bf{X}}_{B}(t)||} \right) \cdot \left( \begin{array}{c} x_{A}(t) - x_{B}(t) \\ y_{A}(t) - y_{B}(t) \end{array} \right) $}\\
    \label{eq:beam} &\textbf{F}_{beam} = k_{beam} \frac{\partial^4}{\partial s^4} \Big( \textbf{X}_C(t) - \textbf{X}^{con} \Big),
\end{align}
where $k_{spr}$ and $k_{beam}$ are the spring and beam stiffnesses, respectively, $R_L(t)$ are the springs resting lengths (set to $ds$, the distance between successive points), $\textbf{X}_{A}=\langle x_A,y_A\rangle$ and $\textbf{X}_{B}=\langle x_B,y_B\rangle$ are Lagrangian nodes tethered by a spring (see Figure \ref{fig:Model_Geometry}). Note that $R_L(t)$ indicates that the resting lengths could be time dependent. $\textbf{X}_C(t)$ is a Lagrangian point on the interior of the jellyfish bell and $\textbf{X}^{con}$ is the corresponding initial (preferred) configuration of that particular Lagrangian point on the jellyfish bell. The spring stiffnesses were large to ensure minimal stretching or compression of the jellyfish bell itself. Through the beam formulation the bell was capable of bending and hence contracting for locomotive purposes. The $4^{th}$-order derivative discretization for the non-invariant beams is given in \cite{BattistaIB2d:2018}.

Lastly, to mimic the subumbrellar and coronal muscles that induce bell contractions, virtual springs were used. These springs dynamically changed their resting lengths over the course of the entire simulation to contract and expand the bell. Rather than tether neighboring points with virtual springs to model the muscles, we tethered points across the jellyfish bell, for all the Lagrangian points that were below the top hemi-ellipse. The deformation force equation does not change from Eq. (\ref{eq:spring}) besides a different $k_{spr}$, which we called $k_{muscle}$, and now used time-dependent spring resting lengths, given by a preset function $R_L(\tilde{t})$. Each simulation defined the period of one complete bell actuation cycle, $T$, based off its specified $f$. Modular arithmetic was used in which we defined $\tilde{t}= t \mod T$, so that $0\leq\tilde{t}\leq T$. As $p$ denoted the fraction of the period that the bell was actively contracted, the overall time the bell was contracting during one cycle was $pT$. Hence the overall expansion time of a cycle was $(1-p)T$. Thus, $R_L(\tilde{t})$ could be mathematically represented as the following:

\begin{equation}
    \label{eq:RL_muscle} R_L(\tilde{t}) = \left\{\begin{array}{cc} 
        2a\cos\left( \frac{\pi \tilde{t} }{2pT} \right) & \ \  \tilde{t}\leq pT  \\
        2a\sin\left( \frac{\pi}{2(1-p)T}\left(\tilde{t}-pT\right) \right) &\ \ pT<\tilde{t}\leq T 
    \end{array}\right.
\end{equation}

By introducing the kinematic control parameter, $p$, and defining $R_L(\tilde{t})$ as above, we were able to modify the kinematic actuation profile in a continuous manner. This allowed us to study variations in swimming performance under perturbations to the actuation kinematics in a concise manner. However, it is worthwhile to note that prolate jellyfish contract on much faster time scales than their expansion phase  \cite{Ford:2000,Colin:2013,Kakani:2013}. Therefore lower values of $p$ more closely depicted prolate jellyfish kinematics. Note that for a simulation of specific parameters $(\mathrm{Re},f,p)$, each actuation cycle's kinematics are identical to all others in that simulation.

\begin{figure}[H]
\centering
\includegraphics[width=0.975\textwidth]{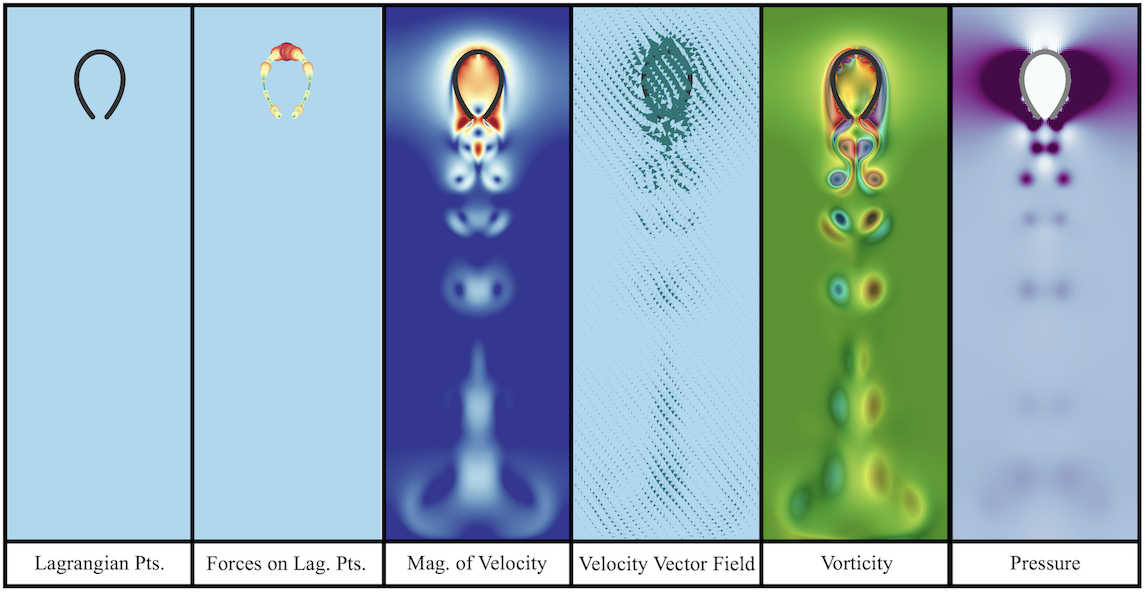}
\caption{A snapshot of a simulation with $(\mathrm{Re},f,p)=(150,0.5,0.5)$ at the peak of contraction during its $5^{th}$ actuation cycle illustrating some of the simulation data obtained at each time-step, e.g., positions of Lagrangian points, magnitude of velocity, the velocity vector field, and vorticity. Note other data not visualized is the fluid pressure and Lagrangian forces spread from the jellyfish onto the Eulerian (fluid) grid.}
\label{fig:example_sim_data}
\end{figure}

Data was stored in $25$ equally spaced time points during each contraction cycle upon running all simulations. The following data was stored:
\begin{enumerate}
    \item Position of Lagrangian Points
    \item Horizontal/Vertical forces on each Lagrangian Point
    \item Fluid Velocity 
    \item Fluid Vorticity
    \item Fluid Pressure
    \item Forces spread onto the Fluid (Eulerian) grid from the Jellyfish (Lagrangian) mesh
\end{enumerate}
We then used the open-source software called VisIt \cite{HPV:VisIt}, created and maintained by Lawrence Livermore National Laboratory for visualization as well as MATLAB \cite{MATLAB:2015a} for post-processing all the simulation data. Figure \ref{fig:example_sim_data} provides a visualization of some of the data produced of a single moment in time for simulation with bell actuation frequency of $f=0.5 s^{-1}$ at a Reynolds Number of $150$ and equal contraction and expansion phase length, $p=0.5$.

We stored 25 time points per contraction cycle in each jellyfish simulation and temporally-averaged data over the 6th-8th contraction cycles to compute three swimming performance metrics - the Strouhal number ($\mathrm{St}$), a non-dimensional forward swimming speed ($1/\mathrm{St}$), and its associated cost of transport ($COT$). First, we found the average swimming speed for a particular simulation, $V_{dim}$, and used this to compute the Strouhal Number, $\mathrm{St}$, 

\begin{equation}
    \label{eq:Str} \mathrm{St} = \frac{fD}{V_{dim}}
\end{equation}
where $f$ is the actuation frequency and $D$ is the maximum bell diameter during an actuation cycle. The above definition of $\mathrm{St}$ used the actuation frequency rather than a vortex shedding frequency, as is commonly done in locomotion studies. Previous studies on animal locomotion have hypothesized that propulsive efficiency is high in a narrow band of $\mathrm{St}$, peaking within the interval $0.2 < St < 0.4$ \cite{Triantafyllou:1991,Taylor:2003,Floryan:2018}. A non-dimensional temporally-averaged forward swimming speed was then defined by taking the inverse of $\mathrm{St}$, i.e., $V=\frac{V_{dim}}{fD}=1/St$. This provided a normalized forward swimming speed based on driving frequency. We then computed the cost of transport ($COT$), which measured the energy (or power) spent per unit speed \cite{Schmidt:1972,Bale:2014,Hamlet:2015}. The dimensional COT, $COT_{dim}$ was defined as

\begin{equation}
    \label{eq:COT_dim} COT_{dim} = \frac{1}{N} \frac{1}{V_{dim}} \displaystyle\sum_{j=1}^N |F_j||U_j|,
\end{equation}
where $F_j$ was the applied contraction force at the $j^{th}$ time step by the jellyfish, $U_{r_{j}}$ was the bell contraction velocity at the $j^{th}$ time step, $N$ was the total number of time steps considered, and $V_{dim}$ was dimensional forward swimming speed during this period of time across all time steps considered. We then non-dimensionalized the cost of transport in the following manner
\begin{equation}
    \label{eq:COT} COT = \frac{COT_{Dim}}{\rho f^2 D^3}.
\end{equation}
The non-dimensional $COT$ calculated here was similar to the energy-consumption coefficient of \cite{Bale:2014}; however, since $D=2a$ and $h$ are conserved across all simulations performed, only one length-scale, $D$, was used in the non-dimensionalization process.

For each simulation performed, the Strouhal number ($\mathrm{St}$), as well as a time-averaged non-dimensional forward swimming speed ($1/\mathrm{St}$) and cost of transport ($COT$) were computed. By performing simulations across highly resolved 2D subspaces of parameters, i.e., ($Re,f$), $(\mathrm{Re},p)$, and $(f,p)$, performance landscapes could effectively be mapped out. Colormap visualizations also offered the ability to qualitatively identify robust parameter subspaces that led to greater swimming performance across each subspace. Once having identified these higher performing subspaces, we sought out to perform global sensitivity analysis in order to find which parameter(s) most significantly affect the swimming performance of prolate jellyfish.

To assess this jellyfish model's overall \textit{global} sensitivity to its parameters, we used Saltelli's extension of the Sobol sequence \cite{Saltelli:2002,Saltelli:2010} to form a discrete subset of parameter values of the overall parameter space in which to simulate the jellyfish. A total of M parameter combinations were selected by this process, $\{(Re,f,p)_j\}_{j=1}^M$, in which to test the model. The same swimming performance metrics ($\mathrm{St}$, $1/\mathrm{St}$, and $COT)$ were computed for each simulation upon finishing, as well as their dimensional analogs, $V_{dim}$ and $COT_{dim}$. Sobol sensitivity analysis was then performed across each performance metric. Although we mostly focused on non-dimensional performance metrics, we computed sensitivities for dimensional quantities, as well. This effort was to determine whether global parameter sensitivities for performance were substantially different across both data presentations, as global sensitivity analysis had not been previously done for jellyfish locomotion models.

Sobol sensitivity analysis is a variance-based sensitivity analysis, which provides the model's overall global sensitivity to parameters, rather than only local sensitivity \cite{Link:2018,Saltelli:2019}. Global sensitivity analysis allowed us to determine which model parameter(s) most significantly caused variations in the model's output across a specified input parameter space. It takes into account the effect of other parameters being varied in conjunction with the one of interest, thus not requiring only one parameter be varied at a time for comparative purposes. On that note, the Sobol sensitivity analysis was able to calculate both first-order parameter sensitivity indices, i.e., perturbations of one parameter at a time, higher-order indices, i.e., those corresponding to perturbations of two or more parameters at a time, and total-order indices, i.e., all combinations of other parameters \cite{Saltelli:2010}. Lastly, the importance of higher-order interactions could be determined by comparing first-order and total-order sensitivity indices. If there are significant differences between these indices, higher-order interactions are present.

\begin{figure}[H]
    \centering
    \includegraphics[width=0.9\textwidth]{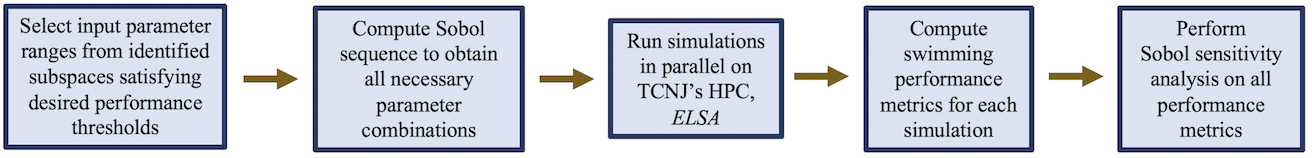}
    \caption{Workflow of the Sobol sensitivity analysis}
    \label{fig:Methods_Workflow}
\end{figure}

Therefore the workflow of the entire Sobol sensitivity analysis process can be summarized as Figure \ref{fig:Methods_Workflow} shows. 

%
%
%
%

\section{Results}

Different modes of locomotion are known to be more effective at certain fluid scales than others \cite{Vogel:1996}. In this work we studied the performance of the jet propulsive technique of a jellyfish with fineness ratio of $1$ across almost four orders of magnitude in $\mathrm{Re}$, within the intermediate Reynolds number regime. This is a particularly interesting (and difficult) regime to study due to the competition and balance of inertial and viscous forces \cite{Klotsa:2019}. As alluded to in past literature, swimming speed is known to be dependent on stroke (in this case, actuation) frequency \cite{Bainbridge:1958,Gray:1968,Steinhausen:2005,Battista:ICB2020}. Moreover, varying the contraction kinematics of prolate jellyfish have shown considerable differences in performance as well \cite{Peng:2012,Alben:2013}. Thus the entire 3D parameter space studied here was $\mathrm{Re}\times f\times p = [0.1,500]\times[0.25,1.25]\times[0.05,0.95].$ 

First, we investigated swimming performance across highly resolved 2D parameter subspaces, corresponding to the following cases:
\begin{itemize}
    \item The ($\mathrm{Re},f$)-subspace for $3$ specific values of the kinematic parameter, $p$
    \item The ($\mathrm{Re},p$)-subspace for $3$ specific frequencies, $f$
    \item The ($f,p$)-subspace for $3$ specific Reynolds numbers, $\mathrm{Re}$.
\end{itemize}
By resolving performance across the above 2D subspaces, any nonlinear effects in performance could be parsed out of the data. Moreover, we were able to identify find robust subspaces in the parameter space which offered greater swimming performance. For this aspect of the study a total of 9,357 fully coupled 2D FSI simulations were performed.

Second, once a subspace of higher performance was identified from the above subspace explorations, we used it to initialize our parameter subspace in order to perform a formal quantitative global sensitivity analysis using Sobol sensitivity analysis. A Sobol sequence was generated using $M=1000$ and $d=3$ (the dimension of the parameter space) using Saltelli's extension of the Sobol sequence \cite{Saltelli:2002,Saltelli:2010}. This gave rise to a total of $M(d+2)=5000$ different parameter combinations that would each need to be simulated in order to find the Sobol sensitivity indices of each performance metric. As global sensitivity analysis had not been previously done on jellyfish locomotion models, we applied it to both non-dimensional and dimensional performance metrics to observe whether there were any discrepancies between the two data forms.

\begin{figure}[H]
    \centering
    \includegraphics[width=0.975\textwidth]{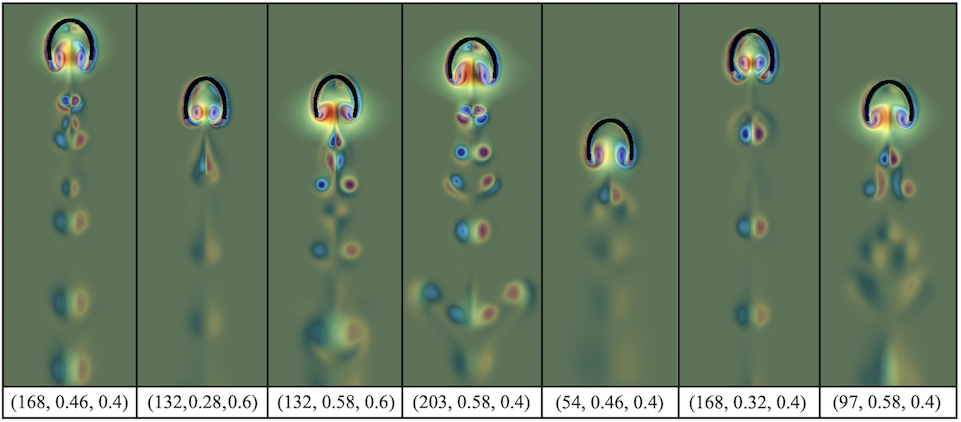}
    \caption{Numerous jellyfish's vortex wake and position after its $5^{th}$ actuation cycle. Different parameter combinations ($\mathrm{Re},f,p$) led to different swimming behavior, both in terms of distance swam and dynamics, as seen by variations in the vortex wakes left behind. The colormap illustrates fluid vorticity.}
    \label{fig:VorticityComparison}
\end{figure}

Hence a grand total of 14,357 fully coupled 2D FSI simulations were performed using The College of New Jersey's high performance computing cluster \cite{TCNJ:ELSA}, requiring just over 1.04 million computational hours. In each simulation as the jellyfish contracts its bell, it expels a vortex ring downward, which by conservation of momentum then propels the jellyfish forward (upwards) \cite{McHenry:2007}. As the bell expands, it draws surrounding fluid up into it. Varying the three parameters of interest $(\mathrm{Re},f,p)$ can lead to substantially different swimming behavior, both in terms of kinematics as well as dynamical performance. Figure \ref{fig:VorticityComparison} shows this through snapshots corresponding to different ($\mathrm{Re},f,p$) cases by giving that case's jellyfish position and fluid vorticity after the $5^{th}$ actuation cycle.  These cases were selected as they all gave rise to pronounced forward swimming, while their vortex wakes were topologically different. We briefly comment that vortex wakes are believed to provide a mechanism by which to understand the functional ecology of a swimming organism \cite{Dabiri:2010c}; however, determining differences in swimming performance based off of vortex wake topology is non-trivial \cite{Floryan:2020}.

The remainder of this section is organized in the following manner. Section \ref{results:2D_Subspaces} explores 2D subspaces for all parameter combinations, i.e., $(\mathrm{Re},f)$, $(\mathrm{Re},p)$, and$(f,p)$ for 3 choices of the third input parameter. Section \ref{results:sensitivity} provides the results for the global sensitivity analysis using Sobol sensitivity analysis. 


%
%

\subsection{Exploring 2D Parameter Subspaces}
\label{results:2D_Subspaces}

Nine total 2D parameter subspaces were explored, each corresponding to a different perpendicular slice from the overall 3D rectangular parameter space, $(\mathrm{Re},f,p)$. Three different cases were performed for each plane: 
\begin{itemize}
    \item $(\mathrm{Re},f)$: $p=0.25, 0.50$, and $0.75$
    \item $(\mathrm{Re},p)$: $f=0.3, 0.5$, and $0.7\ \mathrm{Hz}$
    \item $(f,p)$: $\mathrm{Re}=10, 50$, and $250$.
\end{itemize}

\begin{figure}[H]
    \centering
    \includegraphics[width=0.95\textwidth]{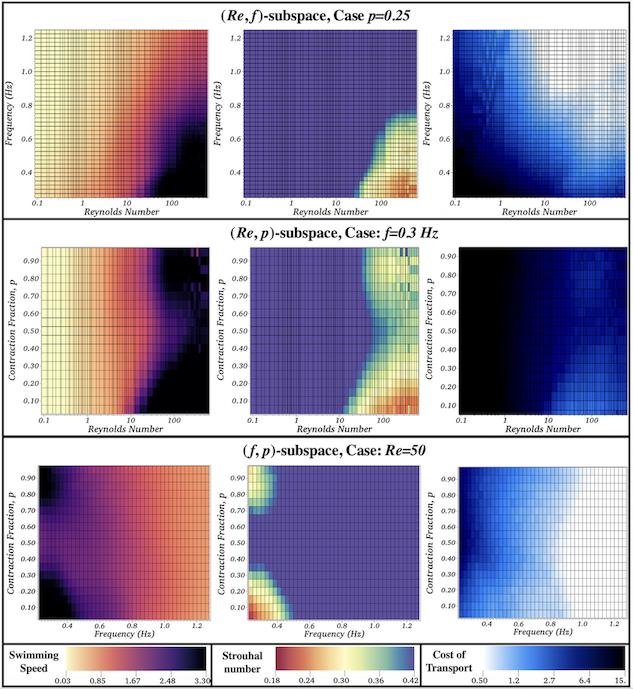}
    \caption{Colormaps illustrating the non-dimensional forward swimming speeds ($1/\mathrm{St}$), Strouhal number ($\mathrm{St}$) and non-dimensional cost of transports ($COT$) over a specific 2D parameter subspace: $(\mathrm{Re},f)$ for the case $p=0.25$ (top row), $(\mathrm{Re},p)$ for the case $f=0.3\ \mathrm{Hz}$ (middle row), and $(f,p)$ for the case of $\mathrm{Re}=50$ (bottom row). Each cell within the colormap represents one particular FSI simulation for the corresponding grid parameter values.}
    \label{fig:All_Colormaps}
\end{figure}

Figure \ref{fig:All_Colormaps} is a summative figure providing data for the three performance metrics of interest - non-dimensional forward swimming speed and cost of transport, as well as Strouhal number in the form of colormaps. The cases illustrated were $(\mathrm{Re},f)$ for $p=0.25$ (top row), $(\mathrm{Re},p)$ for $f=0.3\ \mathrm{Hz}$ (middle row), and $(f,p)$ for $\mathrm{Re}=50$ (bottom row). The colormap values are consistent across every colormap of the same performance metric. From these colormaps parameter regimes in which there is desired swimming performance can be identified. From a quick glance we gathered that non-dimensional swimming speeds $(1/\mathrm{St})$ were maximal for higher $\mathrm{Re}$, lower $f$, and interestingly either lower or higher values of $p$. Values of $p\sim 0.5$ (near the middle of its range) resulted in decreased performance, i.e., lower swimming speed and higher cost of transport. Next we will briefly discuss each 2D parameter subspace separately.

\subsection{Discussing the $(\mathrm{Re},f)$-subspace}
\label{results:ReFreq}

\begin{figure}[H]
\centering
\includegraphics[width=0.925\textwidth]{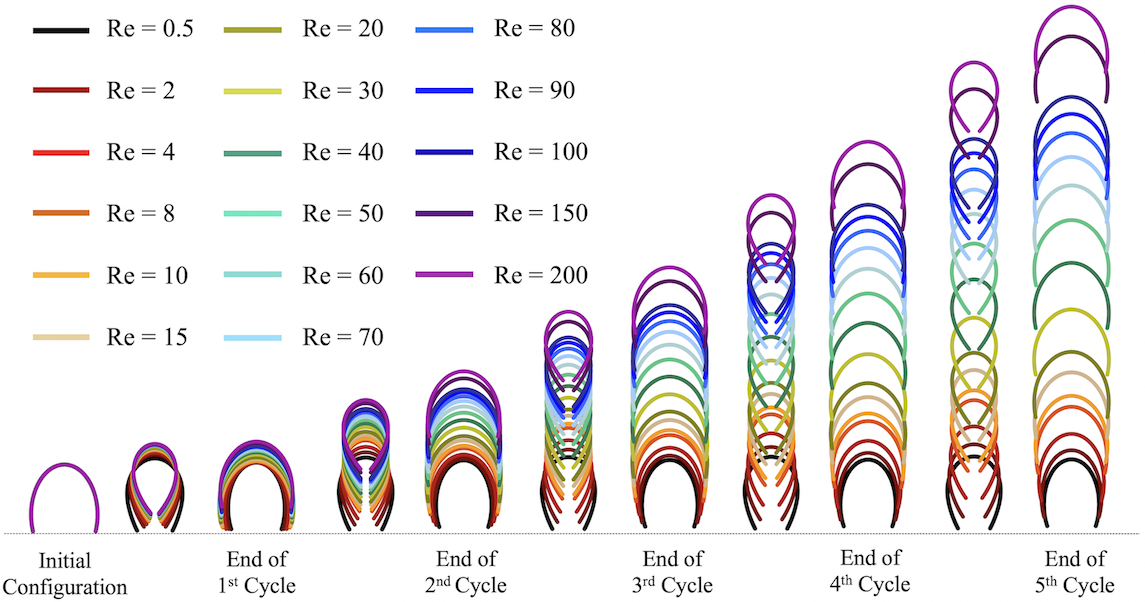}
\caption{Visualization comparing jellyfish swimming for a variety of $\mathrm{Re}$, all with an actuation frequency of $f=0.5\ \mathrm{Hz}$. Forward swimming is more pronounced for larger $\mathrm{Re}$.}
\label{fig:Re_Swim_Compare}
\end{figure}

\begin{figure}[H]
    \centering
    \includegraphics[width=0.95\textwidth]{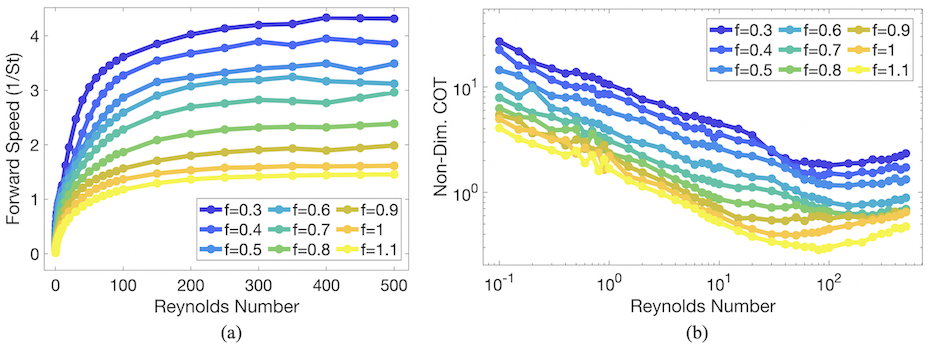}
    \caption{The non-dimensional (a) forward swimming speed and (b) cost of transport as a function of $\mathrm{Re}$ for a variety of different cases involving actuation frequencies, $f$, for $p=0.25$.}
    \label{fig:ReFreq_Data}
\end{figure}

It has been well established that jet propulsion is an effective technique for forward locomotion in jellyfish for fluid scales of $\mathrm{Re}\gtrsim10$ \cite{Hershlag:2011,Yuan:2014,Miles:2019b}. Figure \ref{fig:Re_Swim_Compare} presents snapshots of overlain independent simulations for a variety of $\mathrm{Re}$, but uniform $f=0.5\ \mathrm{Hz}$ and $p=0.5$. As $\mathrm{Re}$ increases, forward swimming appeared to become increasingly more pronounced, see Figure \ref{fig:ReFreq_Data}a. Hoover et al. 2015 \cite{Hoover:2015} observed that in the case of $\mathrm{Re}=150$ and $p=0.5$ there existed an optimal frequency in which to actuate the bell that resulted in maximal (dimensional) swimming speeds that related to resonant properties of the bell itself. However, it has thus far eluded previous studies to vary both $\mathrm{Re}$ and $f$ in observing swimming performance. Here we explored the subspace of $(\mathrm{Re},f)$ for $3$ different values of $p$, $p=0.25, 0.50$, and $0.75$. The data for the subspace corresponding to the case of $p=0.25$ was given in Figure \ref{fig:All_Colormaps}.

The specific $(\mathrm{Re},f)$ subspace provided in Figure \ref{fig:All_Colormaps} illustrates that forward swimming speeds ($1/\mathrm{St}$) were maximal for $f\lesssim 0.75\ \mathrm{Hz}$ and $\mathrm{Re}\gtrsim 50$. Within that same parameter regime, the Strouhal numbers fell within the biologically relevant regime of $0.2<St<0.4$ \cite{Triantafyllou:1991,Taylor:2003,Floryan:2018}. Although the $COT$ was not minimized across those parameter selections, it did slightly overlap with the region in which $COT$ was near minimized, roughly when $f\approx 0.7$ and $100\lesssim Re\lesssim 250$. Interestingly, those parameter ranges also correspond to the maximum and minimum in the \textit{dimensional} swimming speed $(V_{dim})$ and dimensional cost of transport $(COT_{dim})$, respectively, see Figures \ref{fig:app:All_Colormaps_DIM} and \ref{fig:app:All_DIM_Speed}a in Appendix \ref{app_DIM_Data}. Overall, the region of lowest cost of transport corresponds to both higher frequency and higher $\mathrm{Re}$, see Figure \ref{fig:ReFreq_Data}b for a more explicit plot of the data obtained. Figure \ref{fig:ReFreq_Data}a shows that as $\mathrm{Re}$ increases, swimming speed increased but eventually plateaued.  

Previous jellyfish studies have shown the cost of transport for jellyfish is much lower than other metazoans \cite{Gemmell:2013}. However, those studies focused on passive energy recapture in oblate jellyfish, similar to the later computational studies by Hoover et al. 2019 \cite{Hoover:2019}, as the main reason for lower COT in comparison. Nonetheless, this analysis so far has shown the existence of optimal parameter combinations in which jellyfish model exhibits maximal forward swimming speed for minimal cost of transport. Although, across the $(\mathrm{Re},f)$-subspace as a whole, the most robust regions for the fastest forward swimming speeds and minimal cost of transports occur over different frequency ranges.

Different 2D slices of the $(\mathrm{Re},f)$-subspace illustrated dependence on contraction fraction, $p$. Recall that lower values of $p$ corresponded to actuation profiles with shorter contraction times and longer expansion times. The region of maximal forward swimming decreased within the space when $p=0.5$, while it grew larger in both the $p=0.25$ and $p=0.75$ case, see Figure \ref{fig:app:ReFreq_Colormaps} in Appendix \ref{app_ReFreq_Data}. Similarly, the size of the region with biologically relevant $\mathrm{St}$ also decreased for $p=0.50$. By comparing minimal values in the $\mathrm{St}$ panels, it was evident that the fastest forward swimming speeds occurred in the case of $p=0.25$. However, this actuation model showed that it is possible to design fast jetting swimmers that have slow contractions but fast expansions, the opposite of how real prolate jellyfish behave \cite{Ford:2000,Colin:2013,Kakani:2013}.

\subsection{Discussing the $(\mathrm{Re},p)$-subspace}
\label{results:ReP}

Previous studies of jellyfish locomotion have established that swimming performance depends on the kinematics of bell actuation \cite{Peng:2012,Alben:2013} but not how performance may vary across different scales. The middle panels of Figure \ref{fig:All_Colormaps} illustrated that a nonlinear relationship between performance, scale ($\mathrm{Re}$) and the contraction fraction of the overall actuation cycle period, ($p$), across the specific $(\mathrm{Re},p)$ 2D slice corresponding to $f=0.3\ \mathrm{Hz}$. Similar trends in swimming speed ($1/\mathrm{St}$) were observed across $\mathrm{Re}$, such that as $\mathrm{Re}$ increased, speed increased until it eventually began to plateau around $\mathrm{Re}\sim200-300$ regardless of contraction fraction, $p$, for $f=0.3\ \mathrm{Hz}$, see Figure \ref{fig:app:ReP_Speed_COT_f0pt5} in Appendix \ref{app_ReP_Data}. Figure \ref{fig:ReFreq_Data}a showed the same trend as $\mathrm{Re}$ increased, but for varying $f$ and only one $p$, $p=0.25$.

\begin{figure}[H]
    \centering
    \includegraphics[width=0.975\textwidth]{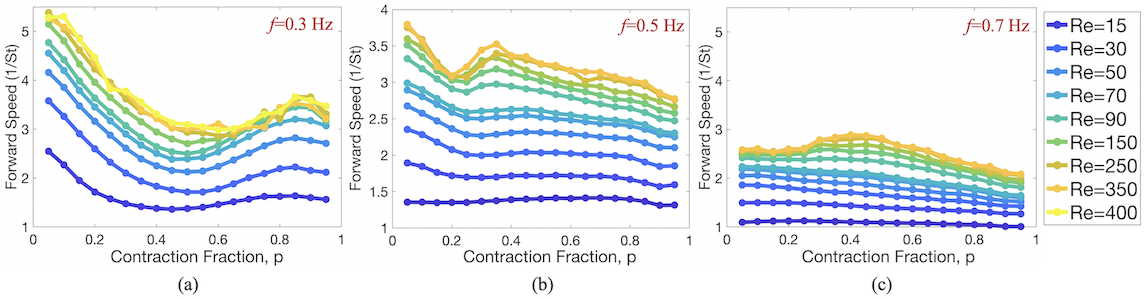}
    \caption{The non-dimensional forward swimming speeds ($1/\mathrm{St}$) as a function $\mathrm{Re}$ for a variety of $p$ for three cases: (a) $f=0.3\ \mathrm{Hz}$, (b) $f=0.5\ \mathrm{Hz}$, and (c) $f=0.7\ \mathrm{Hz}$.}
    \label{fig:ReP_Speed}
\end{figure}

\begin{figure}[H]
    \centering
    \includegraphics[width=0.975\textwidth]{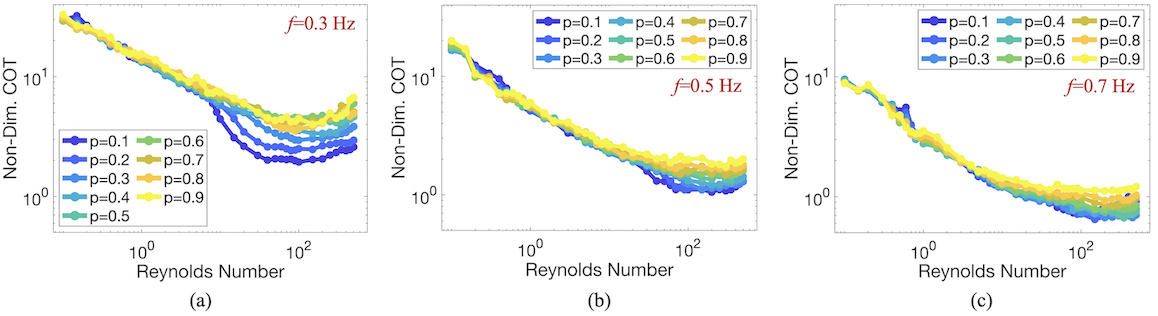}
    \caption{The non-dimensional cost of transport as a function $\mathrm{Re}$ for a variety of $p$ for three cases: (a) $f=0.3\ \mathrm{Hz}$, (b) $f=0.5\ \mathrm{Hz}$, and (c) $f=0.7\ \mathrm{Hz}$.}
    \label{fig:ReP_COT}
\end{figure}

However, as Section \ref{results:ReFreq} briefly alluded, swimming performance has a nonlinear relationship with $p$, see Figure \ref{fig:ReP_Speed}. For $f=0.3\ \mathrm{Hz}$, both higher and lower values of $p$ resulted in faster forward swimming, while $p$ near the middle of its range ($p\sim 0.5$) resulted in a minimum in swimming speed. This same trend was seen in its \textit{dimensional} counterpart data, see Figures \ref{fig:app:All_Colormaps_DIM} and \ref{fig:app:All_DIM_Speed}b in Appendix \ref{app_DIM_Data}. As $f$ was increased, different 2D slices showed substantially different behavior. When $f=0.5\ \mathrm{Hz}$, the minimum in speed was found for $p=0.25$. Moreover, speed increased over $0.25\lesssim p\lesssim0.4$ for $\mathrm{Re}\gtrsim90$, before beginning to decrease again as $p$ increased. When actuation frequency was set at $f=0.7\ \mathrm{Hz}$ a maximum in forward swimming speed appeared for $p\sim0.4$ for $\mathrm{Re}\gtrsim100$. Figure \ref{fig:app:ReP_Colormaps} in Appendix \ref{app_ReP_Data} provides colormaps for all 2D slices investigated across the $(\mathrm{Re},p)$-subspace. Although the $COT$ is highest across the 2D slice corresponding to $f=0.3\ \mathrm{Hz}$, as $\mathrm{Re}$ increases, there is an abrupt transition when $\mathrm{Re}\gtrsim10$ and lower $p$ towards lower $COT$, when compared to all other cases, see Figure \ref{fig:ReP_COT}a. While the $f=0.3\ \mathrm{Hz}$ case resulted in both faster forward swimming and generally higher cost of transport than the other 2D slices that corresponded to higher $f$, the fastest swimming occurred directly over this range of $\mathrm{Re}$ and $p$ in which there was both a sudden drop in and minimal $COT$, but maximal $1/\mathrm{St}$.

\subsection{Discussing the $(f,p)$-subspace}
\label{results:FreqP}

As the previous two sections have illustrated nonlinear relationships amongst swimming performance and $\mathrm{Re}$ and $f$ or $\mathrm{Re}$ and $p$, here we dove into 2D slices of $(f,p)$-subspace, each corresponding to a different $\mathrm{Re}$. Thus, we focused on changing kinematics of the actuation cycle, both in cycle length (frequency) and contraction versus expansion phase lengths within the overall actuation period. This had not been previously studied for jellyfish or other organisms that swim via jet propulsion. The bottom panels of Figure \ref{fig:All_Colormaps} give the performance data for the 2D slice corresponding to $(f,p)$ and $\mathrm{Re}=50$. 

Lower $f$ ($f\lesssim0.4\ \mathrm{Hz}$) and either low or high values or $p$ ($p<0.3$ or $p>0.7$) resulted in higher swimming speeds, see the bottom panels in Figure \ref{fig:All_Colormaps}, corresponding to $(f,p)$ for $\mathrm{Re}=50$. These regions also displayed $\mathrm{St}$ within the biologically relevant range. From minimal values in the $\mathrm{St}$ panel, it was observed that the fastest swimmers over this 2D slice were the ones in which contracted their bells the quickest, i.e., smaller values of $p$. Moreover, this region $(p\lesssim0.3)$ also corresponded to a lower $COT$ than the other region in which produced faster jellyfish, i.e., $p>0.6$. However, the minimal region for $COT$ occurred for $f\gtrsim1.0\ \mathrm{Hz}$ regardless of $p$. As $f$ increased, the swimming speeds converged across multiple of $p$, as Figures \ref{fig:app:FreqP_Colormaps} and \ref{fig:app:FreqP_COT}a in Appendix \ref{app_FreqP_Data} indicated. 

Figure \ref{fig:app:FreqP_Colormaps} provides the data for the other 2D slices explored for cases of $(f,p)$ and $\mathrm{Re}=10,50$, and $250$. As $\mathrm{Re}$ increased, the size of the region in which there were maximal swimming speeds increased, across both $f$ and $p$ directions. However, as the smaller values in the $\mathrm{St}$ panel indicated, the fastest jellyfish were produced for smaller values of $p$.

The \textit{dimensional} form of the data shows the existence of two optimal frequencies to in which to actuate the bell for lower values of $p$, see Figure \ref{fig:app:All_DIM_Speed}c in Appendix \ref{app_DIM_Data}. Hoover et al. 2015 uncovered one of these such frequencies when they showed the existence of an optimum frequency in which to actuate the bell for the specific case of $p=0.50$. That frequency corresponded to the resonant frequency in which to actuate the jellyfish bell to gain a boost in forward swimming speed. Here it corresponded to the frequency of $f\sim 0.75\ \mathrm{Hz}$. Interestingly, our data showed that the emergence of another notable frequency when $p\lesssim0.3$ that appeared to be half that resonant frequency value, $f\sim0.375\ \mathrm{Hz}$. The dimensional data for $\mathrm{St}$ and $COT$ are provided in Figure \ref{fig:app:All_Colormaps_DIM} in Appendix \ref{app_DIM_Data} as well for the case of $\mathrm{Re}=50$ across the $(f,p)$-subspace.

\subsection{Exploring COT vs. Forward Swimming Speed}
\label{results:paramSweeps_Pareto}

\begin{figure}[H]
    \centering
    \includegraphics[width=0.80\textwidth]{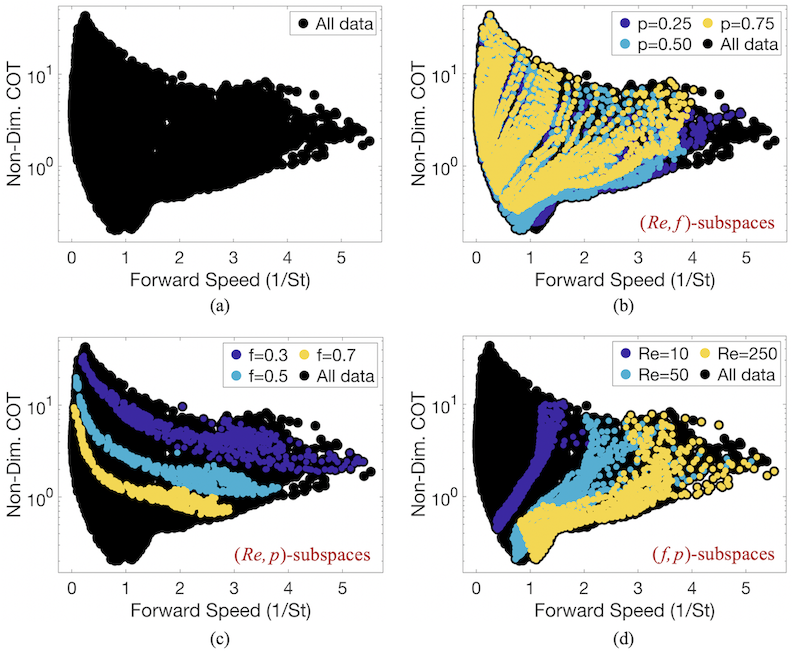}
    \caption{The non-dimensional average cost of transports  and forward swimming speeds ($1/\mathrm{St}$) plotted against each other. (a) Data from all simulations, as well as highlighted regions within the performance spaces corresponding to different ranges of the input parameters: (b) different $\mathrm{Re}$ (c) different $f$, and (d) different $p$.}
    \label{fig:All_ReFreqP_Pareto}
\end{figure}

A Pareto-like front was observed when the non-dimensional cost of transport ($COT$) and the non-dimensional swimming speed ($1/\mathrm{St}$) were plotted against each other \cite{Eloy:2013,Verma:2017,Schuech:2019,Smits:2019}, see Figure \ref{fig:All_ReFreqP_Pareto}a. We deemed these \textit{Pareto-like} since optimal parameter combinations for locomotion would be those in which resulted in maximal swimming speeds for minimal cost of transport, rather than maximal values in both quantities, as how traditional Pareto optimization strategies are constructed. Figure \ref{fig:All_ReFreqP_Pareto}a displays the combination of all the data from Sections \ref{results:ReFreq}-\ref{results:FreqP}. $COT$ was minimal in cases of parameter combinations that resulted in a jellyfish's forward swimming speed of $1/St\sim1$. For different swimming speeds along the curve of minimal $COT$, when speeds decreased from $1$, the $COT$ increased rapidly. For speeds slightly larger than $1$ on the curve of minimal $COT$, there was a sharp increase in $COT$. However, when speeds further increased away from $1$, $COT$ increased much more slowly. As speed increased along the minimal curve from $2$ to $5$, the cost of transport only increased approximately from $0.5$ to $1.0$. 

By probing further into where individual subspaces lied within the overall performance space, patterns emerged, see Figures \ref{fig:All_ReFreqP_Pareto}b-d. For the $(\mathrm{Re},f)$-subspaces, each corresponding to a different $p$, the data mostly all overlapped, see Figure \ref{fig:All_ReFreqP_Pareto}b. That is, given a $p$, different combinations of $(\mathrm{Re},f)$ could almost fill out the entire performance space. A quick glance of the $(\mathrm{Re},p)$ and $(f,p)$-subspaces told a different story. On the $(\mathrm{Re},p)$-subspaces, different values of $f$ placed all $(\mathrm{Re},p)$ combinations into particular regions within the performance space. Higher $f$ corresponded to lower $COT$ for all $3$ slices considered. Although, as $f$ increased, the maximum attainable swimming speeds decreased. Each cluster corresponding to a different $f$ aligned more closely with swimming speed, i.e., along each cluster as speeds increased, the $COT$ did not substantially change. The $(f,p)$-subspaces illustrated that for a given $\mathrm{Re}$, different combinations of $(f,p)$ could result in low or high swimming speeds, with either lower or higher or cost of transports. Smaller $\mathrm{Re}$ had a much more pronounced vertical presence within the performance space, i.e., for the $\mathrm{Re}=10$ case, swimming speeds did not substantially change within the cluster when compared to the higher $\mathrm{Re}$ cases, although its cost of transport substantially varied across the subspace. The analog involving dimensional data for the cost of transport and swimming speeds showed similar clustering trends; the $(\mathrm{Re},f)$-subspace slices all seemed to overlap, while the different cases for the $(\mathrm{Re},p)$ and $(f,p)$-subspaces appeared to cluster more into particular regions within the performance space, see Figure \ref{fig:app_All_DIM_ReFreqP_Pareto} in Appendix \ref{app_DIM_Data}.

%
%

\subsection{Global Sensitivity Analysis}
\label{results:sensitivity}

The jellyfish model's global sensitivity to $(\mathrm{Re},p,f)$ was assessed with Sobol sensitivity analysis. The sensitivity analysis was conducted over the a space in which resulted in substantial forward swimming, i.e., $\mathrm{Re}\times f\times p=[50,250]\times[0.3,1.25]\times[0.05,0.95].$ The $\mathrm{Re}$ were chosen due to the increases in swimming speed observed over this range, before swimming speed ultimately plateaued, see Figures \ref{fig:ReFreq_Data} and \ref{fig:app:ReP_Speed_COT_f0pt5}. The range of $f$ was selected because it centered around the resonant frequency peak of $\sim0.75\ \mathrm{Hz}$ (see Section \ref{results:FreqP}), included lower values of $f$, which resulted in maximal non-dimensional swimming speeds (see Sections \ref{results:ReFreq}-\ref{results:FreqP}, as well as frequencies slightly larger than that of the jellyfish like \textit{Sarsia tubulosa} \cite{Colin:2013} but below that of \textit{Catostylus mosaicus} \cite{Neil:2018b}. Recall that although \textit{Catostylus mosaicus} has a fineness ratio of $\sim1$, its morphology is much more complex, as it includes dense oral arms and an intricate tentacle structure. Thus we decided to keep frequencies within a range for jellyfish without much complex tentacle or oral arm structure, like \textit{Sarsia tubulosa} \cite{Kakani:2013,Katija:2015b}. While it is known that real prolate jellyfish exhibit much shorter contraction phases than expansion phases \cite{Ford:2000,Colin:2013,Kakani:2013}, we elected to study the complete range of $p$ from $0.05<p0.95$ since effective swimming was observed for both ends of the range (see Section \ref{results:FreqP}). 

As mentioned previously, 5000 parameter combinations were sampled within this parameter space, via Sobol sequencing. Each sampled parameter combination was then simulated in order to perform Sobol sensitivty analysis on the desired swimming performance metrics - swimming speed, Strouhal number, and cost of transport. Once they were computed, the Sobol sensitivity indices were calculated \cite{Sobol:2001,Saltelli:2010}. 

\begin{figure}[H]
    \centering
    \includegraphics[width=0.90\textwidth]{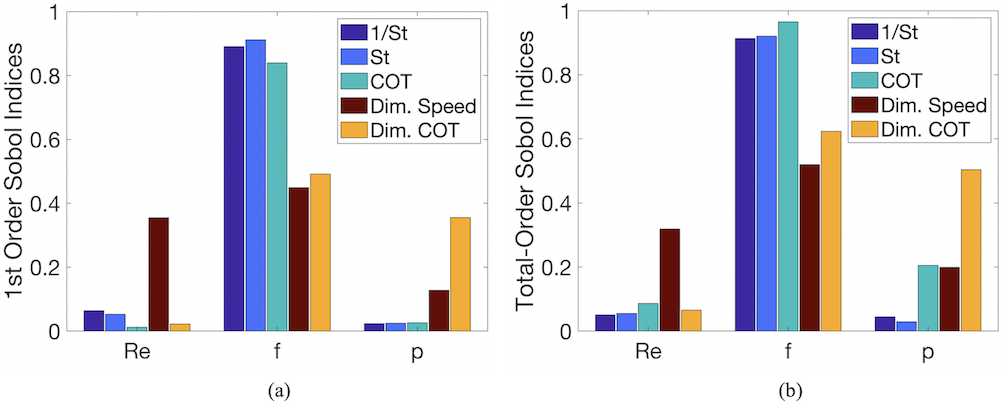}
    \caption{(a) First-order and (b) Total-order Sobol indices corresponding to 3 swimming performance metrics: non-dimensional swimming speed ($1/\mathrm{St}$), Strouhal number ($\mathrm{St}$), non-dimensional cost of transport ($COT$),  for the three input parameters: $\mathrm{Re}$, $f$, and $p$.}
    \label{fig:Sobol_Indices}
\end{figure}

Figure \ref{fig:Sobol_Indices} provides the indices for both the first-order and total-order parameter interactions that were found. Within the restricted parameter space, the swimming performance metrics were most sensitive to variations in actuation (stroke) frequency, $f$. Some higher-order interactions between the parameter exist, as the first-order indices and total-order indices are not equivalent, particularly when comparing the first and total-order indices related to $p$. However, both the first and total-order indices indicated the performance metrics were most sensitive to $f$. This was consistent even for the dimensional analogs of swimming speed and cost of transport. Moreover, the degree of sensitivity to each parameter varied across all performance metrics. Note that while we only varied $\mathrm{Re}, f$, and $p$, it is possible that the efficiency metrics could be more sensitive to other parameters that were not explored here, such as different material properties, i.e., bell stiffnesses \cite{Hoover:2015,Hoover:2019}, duration of inactive (gliding) phases of swimming \cite{Videler:1982,Peng:2012,Alben:2013}, contraction force strength \cite{Hoover:2015,Hoover:2017,Hoover:2019}, or especially the fineness ratio \cite{Peng:2012}, which appears to dictate a jellyfish's main mode of locomotion, either a jet propulsion or jet-padding mode \cite{Ford:2000,Colin:2002,Costello:2008,Weston:2009}.

\begin{figure}[H]
    \centering
    \includegraphics[width=0.90\textwidth]{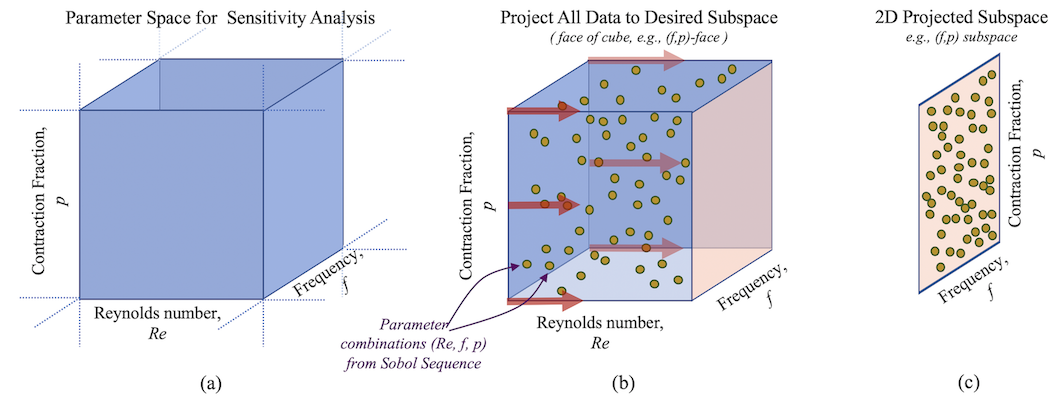}
    \caption{(a) The overall $3D$ parameter space being sampled by Sobol sequences. (b) Illustration of sampling points via Sobol sequences being projected into a $2D$ subspace, here the $(f,p)$ subspace (c) Visualization of where data is projected from a higher dimension space into a $2D$ subspace. Figure adapted courtesy of \cite{Battista:ICB2020b}.}
    \label{fig:Projection}
\end{figure}

The Sobol sensitivity analysis alone granted us sensitivity indices for each parameter; however, it did not divulge in what way the performance metrics were sensitive to each parameter. For example, if $f$ was increased from one value to another, it was unclear whether swimming speeds would increase or decrease. To explore these effects, the entire parameter space sampled via Sobol sequences was projected into three distinct subspaces: $(\mathrm{Re},f)$, $(\mathrm{Re},p)$, and $(f,p)$, as illustrated by the process in Figures \ref{fig:Projection}b-c. Having performed 5000 simulations, the resulting projected subspaces appeared ``filled". We then observed trends within each projected subspace.

\begin{figure}[H]
    \centering
    \includegraphics[width=0.95\textwidth]{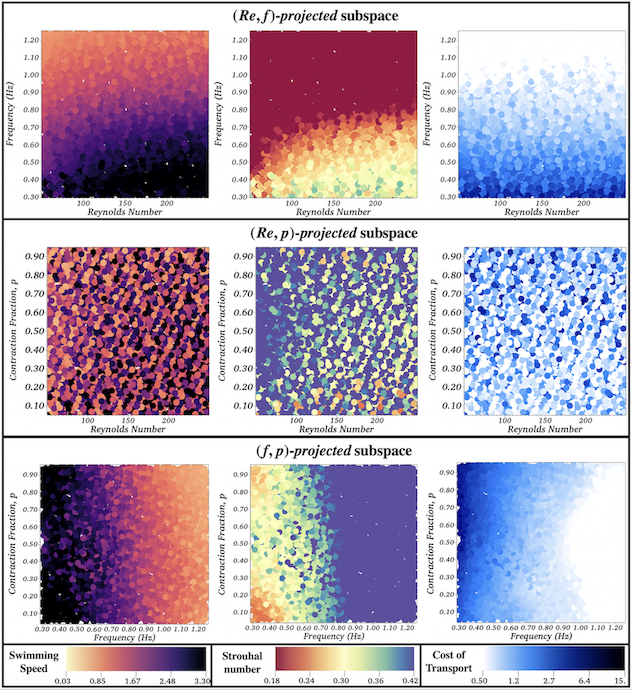}
    \caption{Colormaps of the projected data across all 5000 sampled parameter cases from the Sobol sequence giving the non-dimensional forward swimming speeds ($1/\mathrm{St}$) and cost of transports ($COT$) onto either the $(\mathrm{Re},f)$, $(\mathrm{Re},p)$, or $(f,p)$ subspace.}
    \label{fig:Sobol_Colormap}
\end{figure}

Figure \ref{fig:Sobol_Colormap} provides the performance data across each projected subspace. The gradients in color provided a mechanism to observe emergent patterns within each projected subspace. Qualitatively, across the $(\mathrm{Re},f)$ and $(f,p)$-projected subspaces, the gradient fronts mostly corresponded with variations in $f$. For a given $f$ in these subspaces, some slight dependence was observed on the other parameter, either $\mathrm{Re}$ or $p$, respectively, as the gradient fronts were not completely horizontal in the $(\mathrm{Re},f)$-projected subspaces or vertical in $(f,p)$-projected subspaces. On the other hand, no discernible pattern emerged for any performance metrics within the $(\mathrm{Re},p$)-projected subspace. Thus, qualitative patterns were only noticeable when $f$ was varied, the parameter to whom the model was most sensitive to. Moreover, the most significant variations in these projected subspaces across each performance metric were in the direction of varying $f$. The dimensional analog to Figure \ref{fig:Sobol_Colormap} is provided in Figure \ref{fig:Sobol_DIM_Colormap} in Appendix \ref{app_DIM_Data}.

\begin{figure}[H]
    \centering
    \includegraphics[width=0.825\textwidth]{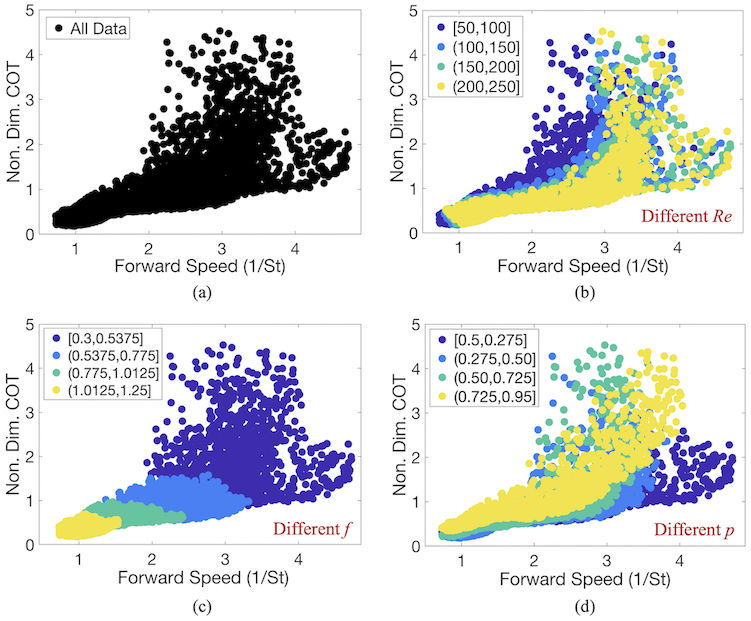}
    \caption{The non-dimensional average cost of transports and forward swimming speeds ($1/\mathrm{St}$) plotted against each other of the data from the Sobol analysis. (a) All data from Sobol simulations, as well as highlighted regions within the performance spaces corresponding to different ranges of the input parameters: (b) different $\mathrm{Re}$ (c) different $f$, and (d) different $p$.}
    \label{fig:Sobol_Pareto}
\end{figure}

Similar to the analysis in Section \ref{results:paramSweeps_Pareto}, a Pareto-like front was observed by plotting the non-dimensional cost of transport ($COT$) against the non-dimensional swimming speed ($1/\mathrm{St}$) \cite{Eloy:2013,Verma:2017,Schuech:2019,Smits:2019}, see Figure \ref{fig:Sobol_Pareto}a. Figures \ref{fig:Sobol_Pareto}b-d, depict how different input parameter ranges fall within the performance space for all the simulations performed for the Sobol sensitivity analysis. Figures \ref{fig:Sobol_Pareto}b and \ref{fig:Sobol_Pareto}d show that for a given $\mathrm{Re}$ or $p$, respectively, that different combinations of the other two input parameters, $(f,p)$ or $(\mathrm{Re},f)$, respectively, could produce a jellyfish whose performance could span the entire performance space. However, different $f$ cluster the performance data into distinct regions within the performance space, see Figure \ref{fig:Sobol_Pareto}b. Higher $f$ pushed the cluster towards both lower swimming speeds and lower cost of transport. This further supported that the notion that this jellyfish model's performance output was most sensitive to the actuation frequency, $f$.

%
%
%
%

\section{Discussion and Conclusion}

Experimental studies of jellyfish locomotion have previously found that not only does bell morphology and scale affect swimming performance \cite{Colin:2002,Dabiri:2005b,Costello:2008} but also the bell kinematics itself \cite{Dabiri:2006,Gemmell:2015}. Here we demonstrated through a 2D fluid-structure interaction model of jellyfish forward locomotion how complex the relationship between scale ($\mathrm{Re}$) and bell kinematics, including actuation frequency ($f$) and the contraction phase percentage of the overall actuation cycle ($p$), is on its achievable swimming performance. Two-dimensional CFD models of jet propulsive prolate jellyfish models have previously shown both qualitative and quantitative agreement with experiments \cite{Sahin:2009b,Hershlag:2011,Alben:2013,Kakani:2013}. Our model recreated similar swimming performance profiles of previous jellyfish locomotion computational models \cite{Hershlag:2011,Alben:2013,Yuan:2014,Hoover:2015}. Notably, we saw similar trends in which forward swimming speed plateaued when $\mathrm{Re}$ was high enough \cite{Hershlag:2011,Miles:2019b}; however, we also determined that this trend was conserved across all contraction frequencies and contraction phase percentages (see Figures \ref{fig:ReFreq_Data} and \ref{fig:app:ReP_Speed_COT_f0pt5}). 

Non-linear relationships arose between the three input parameters $(\mathrm{Re},f,p)$ in which could maximize (or minimize) non-dimensional forward swimming speed ($1/St$) and cost of transport ($COT$) across different 2D slices within the 3D parameter space. Jellyfish with faster contraction times generally resulted in faster forward swimming speeds \cite{Peng:2012,Alben:2013}; however, as Figures \ref{fig:All_Colormaps}, \ref{fig:ReP_Speed}, \ref{fig:app:ReP_Colormaps}, and \ref{fig:app:All_DIM_Speed}c illustrated, given a $f$, maxima in $1/St$ occur for different ranges of $p$. Therefore our model showed that it is not always true that faster swimming results when the contraction time is faster than the expansion time, i.e., $p<0.5$. It also depends on both frequency and scale. Faster contraction times also led to decreased cost of transport within the $\mathrm{Re}$ regimes that also resulted in pronounced swimming, see Figure \ref{fig:ReP_COT}. Peng and Alben 2012 \cite{Peng:2012} had suggested the opposite relationship previously. Furthermore, substantial swimming speeds were also observed for faster expansions than contractions, i.e., $p>0.5$, but these cases also resulted in higher cost of transports. For $\mathrm{Re}>50$, lower $f$ generally resulted in faster forward swimming behavior, see Figures \ref{fig:All_Colormaps} and \ref{fig:app:ReFreq_Colormaps}. For $\mathrm{Re}\gtrsim30$ and a given $p$, an optimal frequency emerged in which resulted in maximized forward swimming in the \textit{dimensional} speed data, see Figures \ref{fig:app:All_Colormaps_DIM} and \ref{fig:app:All_DIM_Speed}, as in Hoover et al. 2015 \cite{Hoover:2015}. The idea of driving jellyfish bells at resonance dates back to 1988 in which the work of Demont and Gosline \cite{Demont:1998c} showed remarkable benefits in both energetic costs and forward propulsion in the prolate jellyfish \textit{Polyorchis penicillatus}. Moreover, a second notable frequency emerged in Figures \ref{fig:app:All_DIM_Speed}a and \ref{fig:app:All_DIM_Speed}c, which maximized forward swimming speed when $\mathrm{Re}\gtrsim30$ and $p\lesssim0.3$. This second frequency appeared to be roughly half that of the optimal (resonant) frequency and also corresponded to minimal cost of transports, see Figure \ref{fig:app:All_Colormaps_DIM}. This second frequency had not been documented previously.

While prolate jellyfish exhibit higher cost of transports than oblate jellyfish \cite{Dabiri:2007,Dabiri:2010b}, which deem them as less efficient swimmers, it is not due to their swimming speeds. They achieve must faster forward swimming speeds but through a more energetically expensive jet propulsive mode of locomotion. Many juvenile jellyfish exhibit jet propulsive modes, due to their higher fineness ratios ($\sim>1$) during development \cite{McHenry:2003,Weston:2009,Blough:2011}. Due to evolutionary constraints of the musculature thickness of cnidarians, the jet propulsive technique becomes ineffective for prolate jellyfish much larger than 10cm \cite{Dabiri:2007,Costello:2008,Colin:2013}. Thus, as they continue to grow and develop, their bell morphologies change to more oblate shapes (lower fineness ratios), which changes their locomotive mode as an adult to rowing or jet-paddling \cite{Weston:2009,Blough:2011}. Theoretical limits for jet propulsion at the interface between oblate and prolate jellyfish were found, see the performance space Figures \ref{fig:All_ReFreqP_Pareto} and \ref{fig:Sobol_Pareto} plotting cost of transport against swimming speed. Clusters within these performance spaces emerged for $\mathrm{Re}$ and $f$ (Figure \ref{fig:All_ReFreqP_Pareto}) or $f$ (Figure \ref{fig:Sobol_Pareto}), to which the latter involved a restricted parameter input space that resulted in greater swimming performance. Furthermore, Pareto-like optimal fronts were attained through this analysis. That is, for a given desired swimming speed within the performance space, parameter combinations could be found in which resulted in the minimal cost of transport.

Global sensitivity analysis showed that the performance metrics were most sensitive to actuation (stroke) frequency across an input parameter space which yield substantial forward swimming behavior, whether in dimensional or non-dimensional units, see Figure \ref{fig:Sobol_Indices}. Although jellyfish swimming performance was known to be dependent on stroke frequency \cite{Peng:2012,Hoover:2015}, the Sobol sensitivity analysis quantitatively suggested that for a jellyfish whose fineness ratio is 1 and that uses a jet propulsive locomotion mode, its swimming performance is \textit{most} affected by stroke frequency. A computational study examining an anguilliform mode of locomotion in nematodes also found that their locomotion mode was most affected by variations in stroke frequency \cite{Battista:ICB2020b}. These analyses were both conducted over a parameter space of high swimming performance, which here included $\mathrm{Re}>50$. Therefore, additional studies are warranted to explore global sensitivity at lower $\mathrm{Re}$, which would include either juvenile \cite{Gordon:2017} or smaller jellyfish \cite{Kakani:2013}. Furthermore, for jellyfish in general, it is not clear how their performance metrics' global sensitivity could change if other input parameters were varied or incorporated, such as the bell's material properties and morphology \cite{Colin:2002,Dabiri:2007,Hoover:2015,Katija:2015,Hoover:2019,Miles:2019b}, contraction kinematics, i.e., jetting or jet-paddling modes \cite{Colin:2002,Dabiri:2007} or gliding times between successive strokes \cite{Videler:1982,Peng:2012,Alben:2013}, contraction strength \cite{Hoover:2015,Hoover:2017,Hoover:2019}, fineness ratio \cite{Peng:2012}. 

Introducing more uncertain model input parameters, such as those quantities suggested above, would have exponentially increased the number of required simulations for the Sobol sensitivity analysis, and hence also the computational expense \cite{Nossent:2011}. Here 5000 simulations were explicitly performed for the sensitivity analysis in order for the Sobol sensitivity indices to converge. These simulations required approximately 360,000 computational hours. Thus, using traditional Sobol sensitivity analysis through the formation of a Sobol sequence is not scalable for fluid-structure interaction models exploring high dimensional input spaces ($>3$). Recently, contemporary methods involving polynomial chaos have emerged as a plausible alternative, as they require far fewer simulations in order for the sensitivity indices to converge \cite{Xiu:2003,Xiu:2003b,Sudret:2008,Waldrop:2018,Waldrop:ICB2020,Waldrop:JRS2020}. Thus they heavily reduce the computational burden; however, by significantly reducing the number of simulations performed may also prohibit the possibility of resolving projected parameter subspaces, like those in Figures \ref{fig:Sobol_Colormap} and \ref{fig:Sobol_DIM_Colormap}. Thereby, while the performance output's global sensitivity would be quantified, it could remain unclear how performance itself varied across the higher dimensional input parameter space.

%
%
%
%

\section*{Acknowledgment}

The authors would like to thank Laura Miller and Alexander Hoover for sharing their knowledge and passion of jellyfish locomotion and Yoshiko Battista for introducing NAB to the world of marine life. We would also like to thank Matthew Mizuhara for his help with sensitivity analyses and Shawn Sivy for his guidance on using the TCNJ HPC more efficiently. We would also like to thank Christina Battista, Robert Booth, Christina Hamlet, Arvind Santhanakrishnan, Emily Slesinger, and Lindsay Waldrop for comments and discussion. J.G.M. was partially funded by the Bonner Community Scholars Program and Innovative Projects in Computational Science Program (NSF DUE \#1356235) at TCNJ. Computational resources were provided by the NSF OAC \#1826915 and the NSF OAC \#1828163. Support for N.A.B. was provided by the TCNJ Support of Scholarly Activity Grant, the TCNJ Department of Mathematics and Statistics, and the TCNJ School of Science.

%
%
%
%

\appendix

%
%

\section{Details on IB}
\label{IB_Appendix}

A two-dimensional formulation of the immersed boundary (IB) method is discussed below. The open-source IB software, \textit{IB2d}, was used to perform all fluid-structure interaction simulations \cite{Battista:2015,BattistaIB2d:2017,BattistaIB2d:2018}. The software itself has been validated \cite{BattistaIB2d:2017} and specific grid resolution convergence tests pertaining to this jellyfish model were also performed previously \cite{BattistaMizuhara:2019}. Additional domain size convergence tests were also performed, with details given below in Appendix \ref{app:conv_check}. For a full review of the immersed boundary method, please see Peskin 2002 \cite{Peskin:2002} or Mittal et al. 2005 \cite{Mittal:2005}. 

%
%

\subsection{Governing Equations of IB}

The conservation of momentum and mass equations that govern an incompressible and viscous fluid are listed below:

\begin{equation} 
   \rho\Big[\frac{\partial\u}{\partial t}({\bf x},t) +\u({\bf x},t)\cdot\nabla \u({\bf x},t)\Big]=  -\nabla p({\bf x},t) + \mu \Delta \u({\bf x},t) + \f({\bf x},t) \label{eq:NS1}
 \end{equation}
  \begin{equation}
      \div \u({\bf x},t) = 0 \label{eq:NSDiv1}
  \end{equation}
where $\u({\bf x},t) $ is the fluid velocity, $p({\bf x},t) $ is the pressure, $\f({\bf x},t) $ is the force per unit area applied to the fluid by the immersed boundary, $\rho$ and $\mu$ are the fluid's density and dynamic viscosity, respectively. The independent variables are the time $t$ and the position ${\bf x}$. The variables $\u, p$, and $\f$ are all written in an Eulerian frame on the fixed Cartesian mesh, $\textbf{x}$. 

The interaction equations, which handle all communication between the fluid (Eulerian) grid and immersed boundary (Lagrangian grid) are the following two integral equations:
\begin{align}
   {\bf f}({\bf x},t) &= \int {\bf F}(s,t)  \delta\left({\bf x} - {\bf X}(s,t)\right) ds \label{eq:force1} \\
   {\bf U}({\bf X}(s,t))  &= \int \u({\bf x},t)  \delta\left({\bf x} - {\bf X}(s,t)\right) d{\bf x} \label{eq:force2}
\end{align}
where ${\bf F}(s,t)$ is the force per unit length applied by the boundary to the fluid as a function of Lagrangian position, $s$, and time, $t$, $\delta({\bf x})$ is a two-dimensional delta function, and ${\bf X}(s,t)$ and ${\bf U}(s,t)$ give the Cartesian coordinates and velocity at time $t$ of the material point labeled by the Lagrangian parameter, $s$, respectively. The Lagrangian forcing term, ${\bf F}(s,t)$, gives the deformation forces along the boundary at the Lagrangian parameter, $s$. Equation (\ref{eq:force1}) applies this force from the immersed boundary to the fluid through the external forcing term in Equation (\ref{eq:NS1}). Equation (\ref{eq:force2}) moves the boundary at the local fluid velocity. This enforces the no-slip condition. Each integral transformation uses a two-dimensional Dirac delta function kernel, $\delta$, to convert Lagrangian variables to Eulerian variables and vice versa.

Using delta functions as the kernel in Eqs.(\ref{eq:force1}-\ref{eq:force2}) is what gives IB its power. To approximate these integrals, discretized (and regularized) delta functions are used. We use the ones given from \cite{Peskin:2002}, i.e., $\delta_h(\mathbf{x})$, 
\begin{equation}
\label{delta_h} \delta_h(\mathbf{x}) = \frac{1}{h^3} \phi\left(\frac{x}{h}\right) \phi\left(\frac{y}{h}\right) \phi\left(\frac{z}{h}\right) ,
\end{equation}
where $\phi(r)$ is defined as
\begin{equation}
\label{delta_phi} \phi(r) = \left\{ \begin{array}{l} \frac{1}{8}(3-2|r|+\sqrt{1+4|r|-4r^2} ),\ \ \ \ \ 0\leq |r| < 1 \\    
\frac{1}{8}(5-2|r|+\sqrt{-7+12|r|-4r^2}),\ 1\leq|r|<2 \\
0 \hspace{1.71in} 2\leq |r|.\\
\end{array}\right.
\end{equation}

%
%

\subsection{Numerical Algorithm}
As stated in the main text, we imposed periodic and no slip boundary conditions on the rectangular domain. To solve Equations (\ref{eq:NS1}), (\ref{eq:NSDiv1}),(\ref{eq:force1}) and (\ref{eq:force2}) we needed to update the fluid's velocity and pressure as well as the position of the boundary and force acting on the boundary at time $n+1$ using data from time $n$. The IB does this in the following steps \cite{Peskin:2002}.

\textbf{Step 1:} Find the force density, ${\bf{F}}^{n}$ on the immersed boundary, from the current boundary configuration, ${\bf{X}}^{n}$.\\
\indent\textbf{Step 2:} Use Equation (\ref{eq:force1}) to spread this boundary force from the Lagrangian boundary mesh to the Eulerian fluid lattice points.\\
\indent\textbf{Step 3:} Solve the Navier-Stokes equations, Equations (\ref{eq:NS1}) and (\ref{eq:NSDiv1}), on the Eulerian grid. Upon doing so, we are updating ${\bf{u}}^{n+1}$ and $p^{n+1}$ from ${\bf{u}}^{n}$, $p^{n}$, and ${\bf{f}}^{n}$. \\
\indent\textbf{Step 4:} Update the material positions, ${\bf{X}}^{n+1}$,  using the local fluid velocities, ${\bf{U}}^{n+1}$, computed from ${\bf{u}}^{n+1}$ and Equation (\ref{eq:force2}).
%

%
%

\section{Computational Grid Width Convergence Check}
\label{app:conv_check}

A convergence test was performed to determine how the width of the computational domain, $L_x$, affected forward swimming speeds and  vortex wake topology. We investigated cases for $Re=\{37.5,75,150,300\}$ and computed forward swimming speed and the subsequent error between cases of different widths of $L_x=\{3,3.5,4,\ldots,8\}$. The height of the domain was fixed with $L_y=12$ and spatial step-sizes was conserved in every computational domain case, i.e., $dx=dy$.

\begin{figure}[H]
\centering
\includegraphics[width=0.975\textwidth]{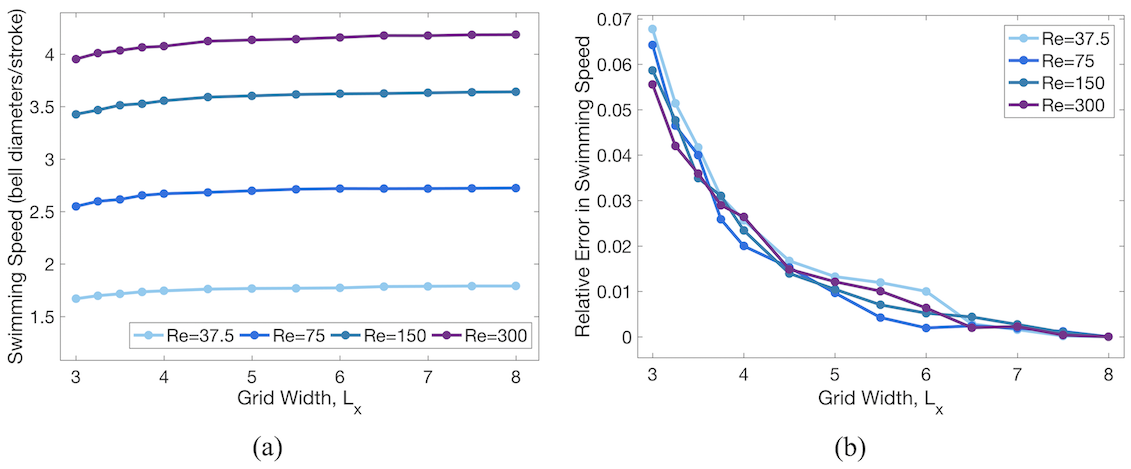}
\caption{$Convergence Check$: $Re=150$ for different computational mesh widths.}
\label{fig:Conv_Check_speed}
\end{figure}

Figure \ref{fig:Conv_Check_speed}a and b provides the swimming speed for every $L_x$ and $\mathrm{Re}$ considered and relative error between each case of $L_x$ against the widest case of $L_x$, respectively. Figure \ref{fig:Conv_Check_speed}a showed that for every case of $\mathrm{Re}$ considered, thinner width simulations produced slightly slower swimming jellyfish; however, the differences in speed were small. As the width gets larger, the relative error decreases, as illustrated by Figure \ref{fig:Conv_Check_speed}b. Moreover, when $L_x=3$, the relative error percentage was $\sim5.5-6.5\%$ and by $L_x=5$, the relative error decreased to $\sim1-1.5\%$. We chose to run simulations using $L_x=5$ in an attempt to minimize computational cost while preserving adequate accuracy. 

\begin{figure}[H]
\centering
\includegraphics[width=0.975\textwidth]{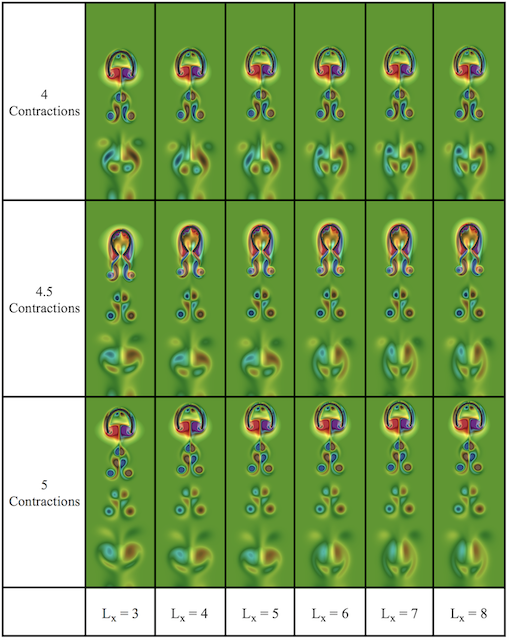}
\caption{\textit{Convergence check} for the case with $(\mathrm{Re},f,p)=(150,0.65,0.5)$ with differing widths of the computational domain, $L_x$.}
\label{fig:Conv_Check_vortices}
\end{figure}

Furthermore, Figure \ref{fig:Conv_Check_vortices} illustrated that qualitative differences were negligible in vortex wake topology in the case of $(\mathrm{Re},f,p)=(150,0.65,0.5)$ across the $4^{th}$ to $5^{th}$ actuation cycle. Cases involving other parameter combinations followed similarly. Subtle differences in vortex dynamics are only observed in down stream vortices when $L_{x}>5$.

%
%

\section{Additional Simulation Data}
\label{app:Additional_Data}

Additional simulation data is provided below. This data complements those presented in the main text of the manuscript. This section is divided into the following sections:
\begin{itemize}
    \item \ref{app_ReFreq_Data}: Data complementing the $(\mathrm{Re},f)$-subspace explorations from Section \ref{results:ReFreq}
    \item \ref{app_ReP_Data}: Data complementing the $(\mathrm{Re},p)$-subspace explorations from Section \ref{results:ReP}
    \item \ref{app_FreqP_Data}: Data complementing the $(f,p)$-subspace explorations from Section \ref{results:FreqP}
    \item \ref{app_DIM_Data}: Swimming performance data in \textit{dimensional} units complementing the non-dimensional data presented in Sections \ref{results:2D_Subspaces}-\ref{results:sensitivity}.
    
\end{itemize}


\subsection{Additional data for the $(\mathrm{Re},f)$-subspace}
\label{app_ReFreq_Data}

\begin{figure}[H]
    \centering
    \includegraphics[width=0.95\textwidth]{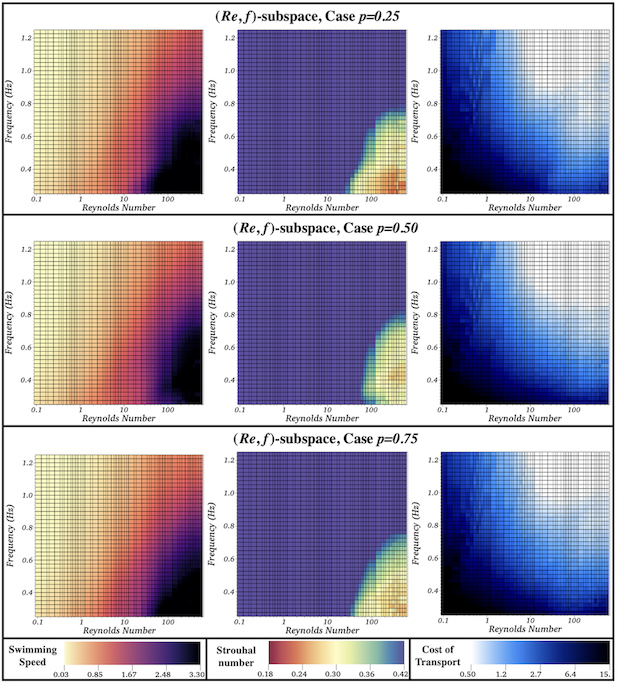}
    \caption{Colormaps corresponding to all cases investigated for the $(\mathrm{Re},f)$-subspace for $p=0.25$ (top row), $p=0.50$ (middle row), and $p=0.75$ (bottom row). Each cell within the colormap represents one particular FSI simulation for the corresponding grid parameter values.}
    \label{fig:app:ReFreq_Colormaps}
\end{figure}

The data corresponding to all 2D slices of the $(\mathrm{Re},f)$ is presented in Figure \ref{fig:app:ReFreq_Colormaps}. The three cases given correspond to the $(\mathrm{Re},f)$-subspace for $p=0.25$ (top row), $p=0.50$ (middle row), and $p=0.75$ (bottom row) in the figure. Variations in the topography of the performance landscapes were minimal across different 2D slices. That is, the sizes of the regions with minimal or maximal performance slightly varied with $p$, but the overall structure remained consistent.

\subsection{Additional data for the $(\mathrm{Re},p)$-subspace}
\label{app_ReP_Data}

\begin{figure}[H]
    \centering
    \includegraphics[width=0.95\textwidth]{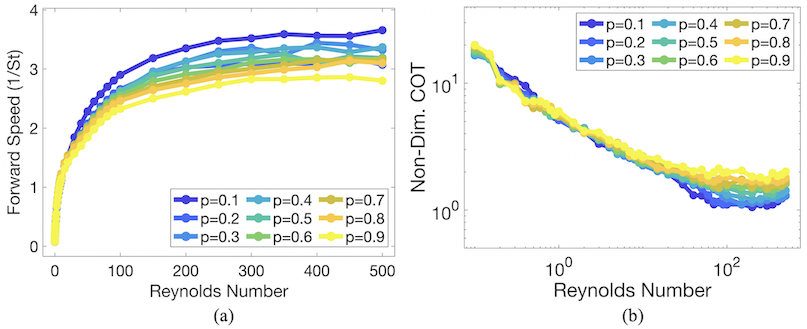}
    \caption{The average non-dimensional (a) forward swimming speed and (b) cost of transport as a function of $\mathrm{Re}$ for a variety of different cases involving contraction fractions, $p$, for the case of $f=0.5\ \mathrm{Hz}$.}
    \label{fig:app:ReP_Speed_COT_f0pt5}
\end{figure}

The average non-dimensional forward swimming speeds ($1/\mathrm{St}$) and cost of transports as a function of $\mathrm{Re}$ are provided in Figure \ref{fig:app:ReP_Speed_COT_f0pt5}, for a variety of different contraction fractions, $p$, and frequency, $f=0.5\ \mathrm{Hz}.$ Faster contraction times (smaller $p$) appeared to correspond to faster jellyfish swimming speeds and lower cost of transports as $\mathrm{Re}$ increased. For $Re\lesssim10$, both swimming speed and cost of transport did not substantially vary for different $p$.  

\begin{figure}[H]
    \centering
    \includegraphics[width=0.95\textwidth]{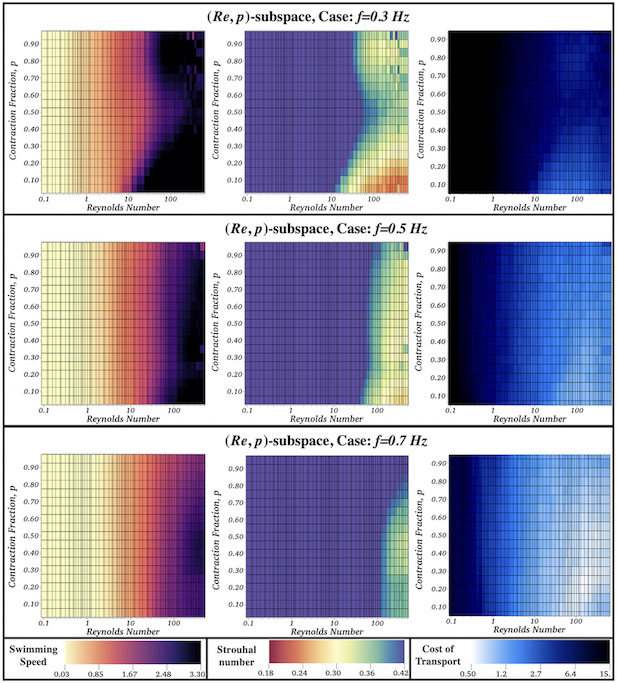}
    \caption{Colormaps corresponding to all cases investigated for the $(\mathrm{Re},p)$-subspace for $f=0.3\ \mathrm{Hz}$ (top row), $f=0.5\ \mathrm{Hz}$ (middle row), and $f=0.7\ \mathrm{Hz}$ (bottom row). Each cell within the colormap represents one particular FSI simulation for the corresponding grid parameter values.}
    \label{fig:app:ReP_Colormaps}
\end{figure}

The data corresponding to all 2D slices of the $(\mathrm{Re},p)$-subspace are given in Figure \ref{fig:app:ReP_Colormaps}. The three cases shown correspond to the $(\mathrm{Re},p)$-subspace for $f=0.3\ \mathrm{Hz}$ (top row), $f=0.5\ \mathrm{Hz}$ (middle row), and $f=0.7\ \mathrm{Hz}$ (bottom row) in the figure. As $f$ varied, there were substantial differences in the topography of each performance landscape. Generally, as $f$ increased the size of the region corresponding to maximal forward swimming speeds decreased, while the region for minimal cost of transport increased in each successive landscape.

\subsection{Additional data for the $(f,p)$-subspace}
\label{app_FreqP_Data}

\begin{figure}[H]
    \centering
    \includegraphics[width=0.975\textwidth]{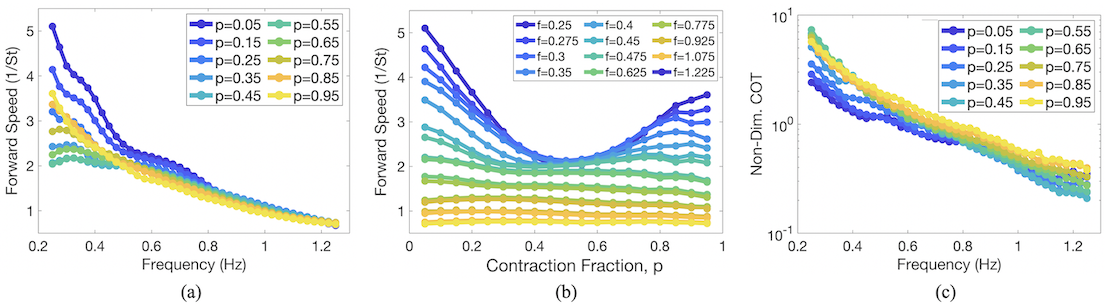}
    \caption{The average non-dimensional swimming speed as (a) a function $f$ for a variety of $p$ and (b) a function of $p$ for a variety of $f$ as well as (c) the non-dimensional cost of transport as a function of $f$ for a variety of $p$.}
    \label{fig:app:FreqP_COT}
\end{figure}

The average non-dimensional forward swimming speeds ($1/\mathrm{St}$) and cost of transports as a function of $f$ for a variety of $p$ and $p$ for a variety of $f$ are provided in Figures \ref{fig:app:FreqP_COT}a and \ref{fig:app:FreqP_COT}b, respectively, for the case of $Re=50$. Swimming speeds all converged for higher frequencies, regardless of contraction fraction, $p$. However, for smaller $f$, minimal values in swimming speed arose for $p\sim 0.5.$ Higher cost of transport was associated with smaller frequencies, see Figure \ref{fig:app:FreqP_COT}c. Moreover, faster contraction times (smaller $p$) resulted in decreased cost of transport.

\begin{figure}[H]
    \centering
    \includegraphics[width=0.95\textwidth]{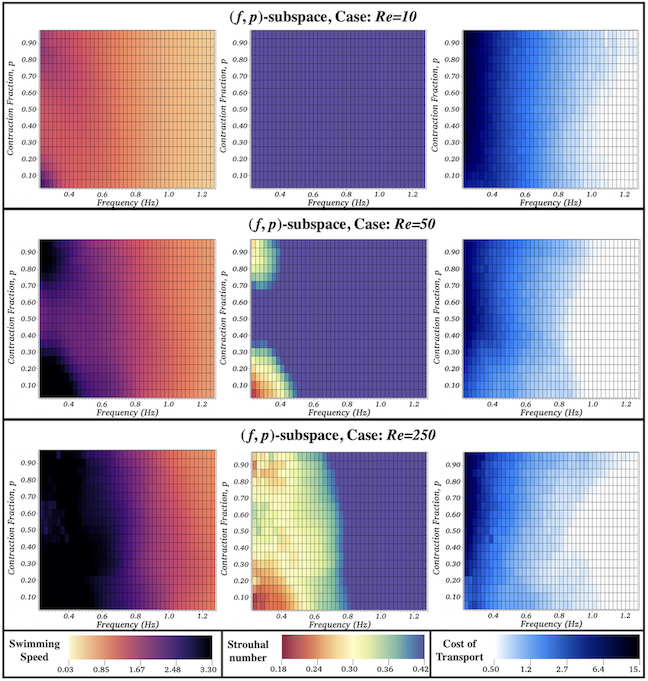}
    \caption{Colormaps corresponding to all cases investigated for the $(f,p)$-subspace for $Re=10$ (top row), $Re=50$ (middle row), and $Re=250$ (bottom row). Each cell within the colormap represents one particular FSI simulation for the corresponding grid parameter values.}
    \label{fig:app:FreqP_Colormaps}
\end{figure}

The data corresponding to all 2D slices of the $(f,p)$-subspace are given in Figure \ref{fig:app:FreqP_Colormaps}. The three cases shown correspond to the $(\mathrm{Re},p)$-subspace for $Re=10$ (top row), $Re=50$ (middle row), and $Re=250$ (bottom row) in the figure. As $\mathrm{Re}$ varied, there were substantial differences in the topography of each performance landscape. Generally, as $\mathrm{Re}$ increased the size of the regions corresponding to maximal forward swimming speeds and minimal cost of transport both increased in each successive landscape. However, these regions did not overlap for any $\mathrm{Re}$ case.

\subsection{Swimming Performance Data in Dimensional Units}
\label{app_DIM_Data}

\begin{figure}[H]
    \centering
    \includegraphics[width=0.95\textwidth]{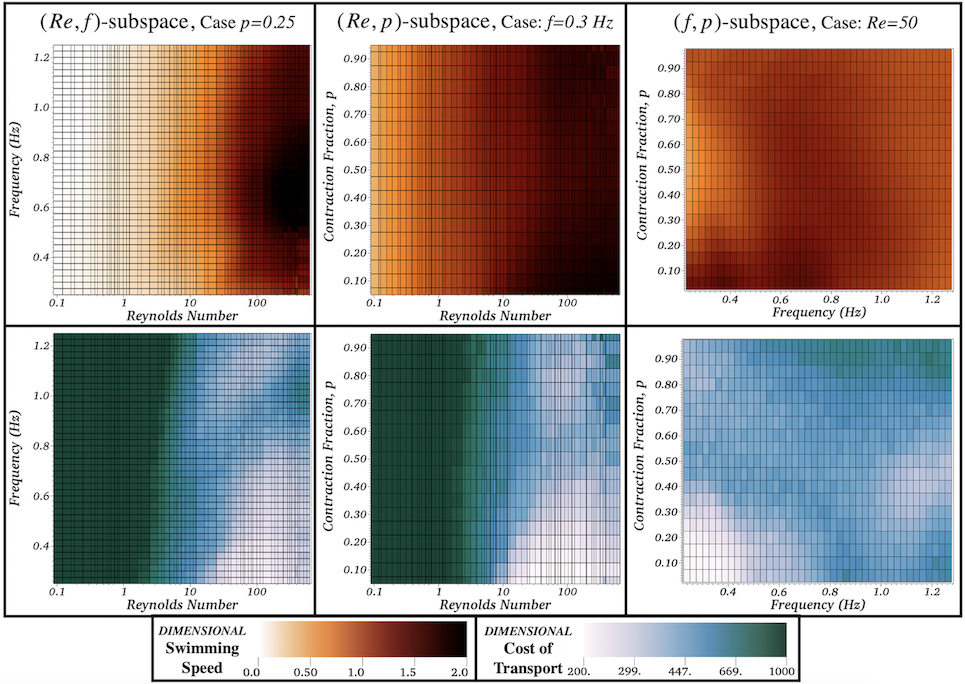}
    \caption{Colormaps illustrating the \textit{dimensional} forward swimming speed ($V_{dim}$) and cost of transport ($COT_{dim}$) to complement the non-dimensional performance data from Figure \ref{fig:All_Colormaps} in Section \ref{results:2D_Subspaces}. Performance data is represented across 2D parameter subspace: $(\mathrm{Re},f)$ for the case $p=0.25$ (left column), $(\mathrm{Re},p)$ for the case $f=0.3\ \mathrm{Hz}$ (middle column), and $(f,p)$ for the case of $Re=50$ (right column). Each cell within the colormap represents one particular FSI simulation for the corresponding grid parameter values.}
    \label{fig:app:All_Colormaps_DIM}
\end{figure}

The \textit{dimensional} data that complements Figure \ref{fig:All_Colormaps} is provided in Figure \ref{fig:app:All_Colormaps_DIM}. The 2D slice corresponding to the $(\mathrm{Re},f)$-subspace for $p=0.25$ showed the existence of a robust region of maximal swimming speeds for $f\sim0.7\ \mathrm{Hz}$ and $Re\gtrsim100$. Similarly, the $(\mathrm{Re},p)$-subspace for $f=0.3\ \mathrm{Hz}$ showed maximal swimming speeds for $Re\gtrsim50$ and either low $p$ (fast contraction times) or high $p$ (slow contraction times). Minimal swimming speeds occurred when contraction and expansion times were approximately equal ($p\sim0.5$). The $(f,p)$-subspace for $Re=50$ divulged maximal swimming speed clusters around $f\sim0.35\ \mathrm{Hz}$ and $f\sim0.7\ \mathrm{Hz}$ for faster contraction times ($p\lesssim0.2$). However, other substantially fast regions appeared over different $(f,p)$ ranges, e.g., $0.6\gtrsim f\gtrsim0.8$ and $0.20\gtrsim p\gtrsim 0.70$. Lower costs of the transport were generally observed in the $(\mathrm{Re},f)$- and $(\mathrm{Re},p)$-subspaces as $\mathrm{Re}$ increased. In those subspaces for $Re\gtrsim50$, nonlinear dependence on $f$ and $p$, respectively, led to regions of minimal cost of transport. The $(f,p)$-subspace illustrated distinct parameter clusters in which led to minimal cost of transport. The region encasing $p\lesssim0.2$ and $0.3\gtrsim f\gtrsim0.4$ had some of the fastest swimming speeds and lowest costs of transport within the subspace.

\begin{figure}[H]
    \centering
    \includegraphics[width=0.95\textwidth]{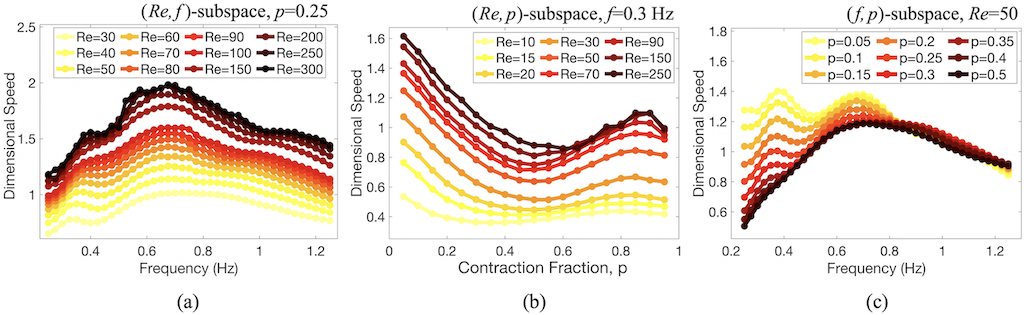}
    \caption{\textit{Dimensional} forward swimming speed ($V_{dim}$) provided for different 2D parameter subspaces.}
    \label{fig:app:All_DIM_Speed}
\end{figure}

Explicit plots of the data complementing the colormaps in Figure \ref{fig:app:All_Colormaps_DIM} are provided in Figure \ref{fig:app:All_DIM_Speed}. Nonlinear relationships emerged across every 2D subspace explored. The optimal actuation frequency found in Figure \ref{fig:app:All_DIM_Speed}a for the case of $Re=150$ agreed with the resonant frequency analysis from Hoover et al. 2015 \cite{Hoover:2015}.

\begin{figure}[H]
    \centering
    \includegraphics[width=0.80\textwidth]{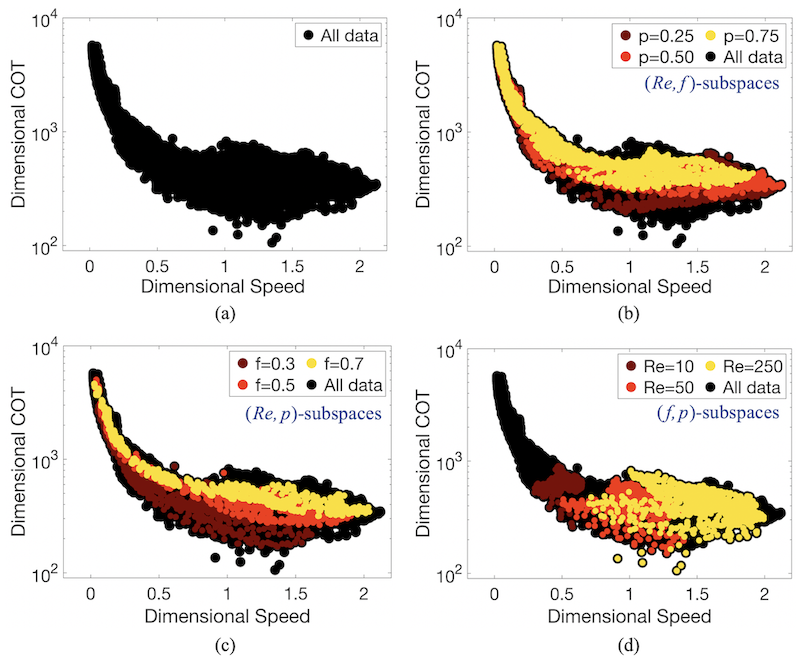}
    \caption{The \textit{dimensional} average cost of transports and forward swimming speeds ($1/\mathrm{St}$) plotted against each other. (a) Data from all simulations, as well as highlighting regions within the performance landscape corresponding to different ranges of the input parameters: (b) different $\mathrm{Re}$ (c) different $f$, and (d) different $p$.}
    \label{fig:app_All_DIM_ReFreqP_Pareto}
\end{figure}

The dimensional cost of transport was also plotted against the dimensional forward swimming speed, see Figure \ref{fig:app_All_DIM_ReFreqP_Pareto}. All of the simulation data corresponding to every 2D subspace study, i.e., Sections \ref{results:ReFreq}-\ref{results:FreqP}, is given in Figure \ref{fig:app_All_DIM_ReFreqP_Pareto}a, whiles Figure \ref{fig:app_All_DIM_ReFreqP_Pareto}b-d illustrate where different parameters lies across the performance space. Generally Figures \ref{fig:app_All_DIM_ReFreqP_Pareto}c and \ref{fig:app_All_DIM_ReFreqP_Pareto}d have distinct regions where certain data lies within the performance space for different frequencies and $\mathrm{Re}$, respectively. Lower $f$ generally resulted in lower cost of transport. Higher $\mathrm{Re}$ generally resulted in higher swimming speeds. However, both Figures \ref{fig:app_All_DIM_ReFreqP_Pareto}b and \ref{fig:app_All_DIM_ReFreqP_Pareto}c displayed that given a $p$ or $f$, combinations of the other two parameters, $(\mathrm{Re},f)$ or $(\mathrm{Re},p)$, respectively, could produce a jellyfish with any forward swimming speed in the performance space.. 

\begin{figure}[H]
    \centering
    \includegraphics[width=0.95\textwidth]{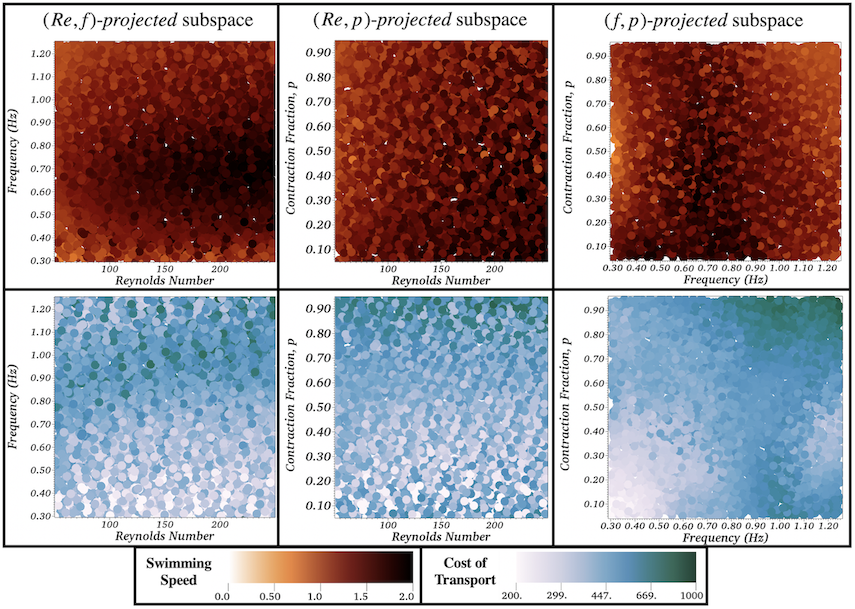}
    \caption{Colormaps of the projected data across all 5000 sampled parameter cases from the Sobol sequence giving the \textit{dimensional} forward swimming speeds and cost of transports onto either the $(\mathrm{Re},f)$, $(\mathrm{Re},p)$, or $(f,p)$ subspace.}
    \label{fig:Sobol_DIM_Colormap}
\end{figure}

The \textit{dimensional} data that complements Figure \ref{fig:Sobol_Colormap} is provided in Figure \ref{fig:Sobol_DIM_Colormap}. The colormaps illustrate data for all 5000 sampled parameter combinations from the Sobol sensitivity analysis when projected onto the $(\mathrm{Re},f)$-subspaces (left column), $(\mathrm{Re},p$)-subspaces (middle column), or $(f,p)$-subspaces (right column), for both forward swimming speed and cost of transport. Patterns emerged within $(\mathrm{Re},f)$ and $(f,p)$ projected subspaces giving swimming speeds, where the trends aligned mostly with varying values of frequency, $f$. That is, variations in frequency led to the most significant overall changes in swimming speed. Variations in the cost of transport within the $(\mathrm{Re},f)$ projected subspace most aligned with changing frequency, as well. However, more nonlinear relationships emerged in the cost of transport within the $(f,p)$ projected subspace. Across the $(\mathrm{Re},p)$ projected subspace no discernible patterns emerged in either swimming speed or cost of transport.

%
%

\bibliographystyle{spmpsci}
\bibliography{jelly}

\end{document}